\documentclass[twocolumn,showpacs,groupedaddress,assymb,amsmath]{revtex4}
\usepackage{}
\usepackage{graphicx}
\usepackage{amssymb}
\usepackage{amsmath}

\begin{document}

\title{Exact non-Markovian master equation for a driven damped  two-level system}
\author{H. Z. Shen$^{1}$, M. Qin$^{1}$, Xiao-Ming Xiu$^{1,2}$, and X. X. Yi$^3$
\footnote{Corresponding address: yixx@nenu.edu.cn}}
\affiliation{$^1$School of Physics and Optoelectronic Technology\\
Dalian University of Technology, Dalian 116024 China\\
$^2$Department of Physics, College of Mathematics and\\
Physics, Bohai University, Jinzhou 121013, China\\
$^3$ Center for Quantum Sciences and School of Physics, Northeast
Normal University, Changchun 130024, China }
\date{\today}

\begin{abstract}
Driven two-level system  is a useful model to describe many quantum
objects, particularly in quantum information processing. However,
the exact master equation for such a system is barely explored.
Making use of the Feynman-Vernon influence functional theory, we
derive an exact non-Markovian master equation for the driven
two-level system and  show the lost feature in the perturbative
treatment for this system. The perturbative treatment leads to the
time-convolutionless (TCL) and the Nakajima-Zwanzig (NZ) master
equations. So to this end, we derive the time-convolutionless (TCL)
and the Nakajima-Zwanzig (NZ) master equations for the system and
compare the dynamics given by the three master equations. We find
the validity condition for the TCL and NZ master equations. Based on
the exact non-Markovian master equation, we analyze   the regime of
validity for the secular approximation in the time-convolutionless
master equation and discuss the leading corrections of the
nonsecular terms to the quantum dynamics, significant effects are
found in the dynamics of the driven system.
\end{abstract}

\maketitle
\section{Introduction}
The  dynamics of open quantum systems
\cite{Alicki2007717,Breuer2002,Weiss2008} has attracted much
attention and becomes active again in recent years due to its
possible applications in quantum information science
\cite{DiVincenzo1998393,Knill2001409,Cirac199959,DiVincenzo200048,Cirac199778,
Duan200367}. Indeed the study of coupled system-environment system
is an long standing endeavor in many fields of  physics including
quantum optics
\cite{Walls1994,Scully1997,Gardiner2000,Carmichael1993}, atomic
optics \cite
{Mandel1995,Weissbluth1988,Vogel1994,Compagno1995,Carmichael1993}
and condensed matter physics
\cite{Caldeiral1983149,Leggett198759,Weiss2008}. The coupling  of
the system to its environment leads to dissipation and dephasing
with flows of energy or information from the system to the
environment \cite{Breuer2002,Weiss2008}. The back flowing of
information from the environment to the system determines the
Markovianity of the dynamics.

Driven two-level model is available to effectively describe many
actual physical systems, for example, a quantum bit in quantum
information processing. Thus the theoretical analysis as well as the
practical implementation of the driven two-level systems brings us a
renewed topic. There are several ways to create a driven two-level
system (or qubit) today by current quantum technologies, each
exploits different approaches or in different quantum systems. For
instance, by means of quantum optics and in microscopic quantum
objects (electrons, ions, atoms) in traps, quantum dots, and quantum
circuits
\cite{Majer200594,Berkley2003300,Pashkin2003421,Bellomo200799}.
Different implementations of qubit \cite{Nielsen2000,Barenco199574}
are subjected to different types of environmental noise
\cite{Biercuk2009458}, most environments are assumed Markovian
\cite{Das200942,Sinaysky200878} and the dynamics of system was
studied perturbatively in the literatures.

In recent years, an increasing interest has been paid to developing
a non-Markovian generalization for open quantum system theory, some
of them are formulated in terms of non-local   time evolutions.
There exist diverse formalisms for describing memory effects,
including the generalization of the Lindblad master equation from
time-independent dissipative rates to time-convoluted kernel
functions. A wide class of both phenomenological and theoretical
approaches were formulated for building and characterizing this type
of master equations, which in turn lead to a completely positive
map.

By means of the Feynman-Vernon influence  functional theory
\cite{Feynamn196324118,Caldeira1983121587,Hu1992452843,zhang199062867,
Anastopoulos200062},  exact master equations describing the general
non-Markovian dynamics of a wide range of open quantum system have
been recently developed, e.g. quantum Brownian motion
\cite{Hu1992452843,Karrlein199755153,Haake198532}, single-mode
cavity \cite{An20093241737} and two entangled cavities
\cite{An200776042127,An200990317} with vacuum fluctuations, \textbf{spin-boson model}
\cite{Lucke199911110843}, coupled
harmonic oscillators
\cite{Chow200877011112,Paz2008100220401,Paz200979032102}, quantum
dot in nanostructures \cite{Tu200878235311,Tu20098631}, various
nanodevices with time-dependent external control field
\cite{Jin201012083013}, nanocavity systems including initial
system-reservoir correlations \cite{Tan201183}, and photonic
networks imbedded in photonic crystals
\cite{Lei20123271408,zhang2012109170402}. However, an exact master
equation for driven systems are very rare.

Projection operator technique is other mean to study the open
quantum system, both the time-convolutionless (TCL)
\cite{Shibata197717,Prataviera201387,Chaturvedi197935} master
equation and the
Nakajima-Zwanzig (NZ) \cite{Nakajima195820,Zwanzig196033,zhang201387032117} master
equation can be derived by this approach. The NZ approach provides
us with a generalized master equation in which the time derivative
of the density operator is connected to the past of the reduced
density matrix through the convolution of the density operator and
an appropriate integral kernel. While the TCL approach leads to a
generalized master equation which is local in time. It seems that
the NZ should run better than the TCL approach in describing the
 non-Markovian effect, since it takes into account the
history of the reduced density matrix. However this is not the case
as we will show later, examples in
\cite{Breuer2002,Breuer1999591633,Yan1998582721,Ferraro200980042112,Xu20011143868,
Schroder2006124084903,Liu200776022312,Haikka201081052103,Haikka2010014047}
confirm this point, namely, the exact dynamics of the open system
can be described via a master equation with time-dependent decay
rate, as in the well-known case of the Hu-Paz-Zhang generalized
master equation\cite{Breuer2002,Hu1992452843}.

In the weak coupling limit,  the non-Markovian master equation for a
driven two-level system coupled to a bosonic reservoir at zero
temperature  has been derived and discussed in
Ref.\cite{Haikka201081052103}. This derivation treat the
system-environment coupling perturbatively, and hence it is
available for weak system-environment couplings. In this paper,
exploiting the Feynman-Vernon influence theory in the coherent state
path integral formalism, we derive an exact non-Markovian master
equation for the driven two-level system. The Feynman-Vernon
influence theory enables us to treat the environment-system coupling
non-perturbatively. The dynamics of the driven open two-level
system, going beyond the TCL, NZ, and Markovian approximations, is
governed by an effective action associated with the influence
functional containing all the influences of the environment on the
system. The  exact master equation is available to examine the
validity of those perturbative approaches applied to  the TCL and NZ
techniques. We show that the TCL approach works better than the NZ
one, since the latter does not guarantee the positivity of the
density matrix when the correlations in the reservoir become strong,
while the former is available for a wider range of values of
reservoir memory time.

The remainder of the paper is organized as follows. In Sec. {\rm
II}, we introduce a model to describe a driven two-level system
subject to reservoir and give a detailed derivation of the influence
functional for the model in the coherent state representation. In
Sec. {\rm III}, an exact non-Markovian master equation describing
the evolution of the driven open two-level system is derived. In
Sec. {\rm IV}, a derivation of the second-order NZ master equation
is presented and the characteristics of the second-order TCL derived
in Ref. \cite{Breuer2002} are discussed, and then we give a
comparison among the exact, TCL, and NZ master equations. In Sec.
{\rm V}, we investigate the validity of the secular approximation in
Markovian and non-Markovian regimes, respectively. Discussions and
conclusions are given in Sec. {\rm VI}.
\section{atomic coherent state path-integrate approach to the driven open two-level system}
\subsection{Model Hamiltonian}
\textbf{We start by considering  a two-level system with Rabi
frequency $\omega _0$ driven by an external laser of frequency
$\omega_L$. The two-level atom is embedded in a bosonic reservoir at
zero-temperature modeled by a set of  infinite   harmonic
oscillators. In a rotating frame,  the Hamiltonian of such a system
(system plus environment) takes}
\begin{eqnarray}
H = {H_S} + {H_E} + {H_I},\label{hHh12}
\end{eqnarray}
with
\begin{equation}
\begin{aligned}
{H_S} =& \Delta  \cdot {\sigma _ + }\sigma_-  + \Omega {\sigma _ x },\\
{H_E} =& \sum\limits_k {{\Omega _k}a_k^\dag {a_k}} ,\\
{H_I} =& \sum\limits_k {{g_k}{\sigma _ + }{a_k} + H.c.} ,
\label{H}
\end{aligned}
\end{equation}
\textbf{where $\Delta  = {\omega _0} - {\omega _L}$,  ${\Omega _k} =
{\omega _k} - {\omega _L}$, and ${\sigma _x} = {\sigma _ +}
+\sigma_-$. $\Omega$ is the driven strength, and H.c. stands for the
Hermitian conjugation. $\sigma  _+= \left| e \right\rangle
\left\langle g \right|$ is the Pauli matrix. ${a_k}$ and ${g_k}$ are
the annihilation operator and   coupling constants, respectively. In
the following we shall start with this Hamiltonian (\ref{hHh12}) and
derive all master equations in this paper.}

\subsection{Coherent state representation}
The starting point of analysis is to observe that the lowing and
raising operators of the atomic transition operators ${\sigma _ + }
= \left| e \right\rangle \left\langle g \right|$ and $\sigma_-  =
\left| g \right\rangle \left\langle e \right|$ satisfy
anticommutation rules similar to those of   fermions, i.e.,
\begin{equation}
\begin{aligned}
\left\{ {\sigma_-,{\sigma _ + }} \right\}  =&
\left| e \right\rangle  \left\langle e \right| + \left| g \right\rangle \left\langle g \right| \equiv 1,\\
\left\{ {\sigma_- ,\sigma_-} \right\} = & \left\{ {{\sigma _ +
},{\sigma _ + }} \right\} = 0, \label{antio}
\end{aligned}
\end{equation}
where $\{ A,B\}  = AB + BA$. Identifying the ground  state $\left| g
\right\rangle$ with the fermionic vacuum,  we can therefore treat
${\sigma _ + }$ and $\sigma_-$ as fermionic creation and
annihilation operators, respectively. Following Ref.
\cite{Cahill1999591538}, we  introduce a couple of conjugate
Grassmann variables $\zeta $ and ${\bar{\zeta}}$ imposing standard
anticorrelation with the annihilation and creation operators of the
system.

Therefore, coherent states are defined as a tensor product of
states generated by exponentiated operation of a creation operator
and a suitable label on a chosen fiducial
state \cite{Glauber1963131,Shresta200571022109,
Anastopoulos200062,zhang199062867,Ghosh201286011138}
\begin{eqnarray}
\left| \mathbf{z} \right\rangle  =  \prod\limits_k {\left| {{z_k}}
\right\rangle } ,\left| {{z_k}} \right\rangle  = \exp (a_k^\dag
{z_k})\left| {{0_k}} \right\rangle , \label{coherentstate1}
\end{eqnarray}
and
\begin{eqnarray}
\left| \zeta  \right\rangle  = \exp ({\sigma _ + }\zeta )\left| g \right\rangle.
\label{coherentstate2}
\end{eqnarray}
For bosonic coherent states  defined in Eq.~(\ref{coherentstate1}),
the label ${z_k}$ is a complex number, and for atomic coherent
states defined in Eq.~(\ref{coherentstate2}), the label $\zeta $ is
a Grassmannian or anticommuting number. A state of the combined
atom-field system can be expanded in a direct product of coherent
state
\begin{eqnarray}
\left| \mathbf{z}\zeta \right\rangle = \left| \mathbf{z}
\right\rangle  \otimes \left| \zeta  \right\rangle.
\end{eqnarray}
Atomic and bosonic coherent states possess  the well-known
properties  such as being nonorthogonal
\begin{eqnarray}
\begin{aligned}
\left\langle {\mathbf{z}}
 \mathrel{\left | {\vphantom {z {z'}}}
 \right. \kern-\nulldelimiterspace}
 {{\mathbf{z'}}} \right\rangle  =  \exp (\sum\limits_k  {\bar{z}_k {z'_k}} ),\left\langle {\zeta }
 \mathrel{\left | {\vphantom {\zeta  {\zeta '}}}
 \right. \kern-\nulldelimiterspace}
 {{\zeta '}} \right\rangle  = \exp ({\bar{\zeta} }\zeta '),\\
  \end{aligned}
\end{eqnarray}
\begin{eqnarray}
\begin{aligned}
{a_k}\left| {{z_k}} \right\rangle  = {z_k}\left| {{z_k}} \right\rangle ,\sigma_-\left| \zeta  \right\rangle  = \zeta \left| \zeta  \right\rangle ,
 \label{ortho2}
 \end{aligned}
\end{eqnarray}
where $\bar{z}_k$ and $\bar{\zeta}$ denote the conjugation of $z_k$
and $\zeta$, respectively. Despite their nonorthogonality, both
types of coherent states form an over-complete basis set
\begin{eqnarray}
\int {d\varphi ({\mathbf{z}})\left| \mathbf{z} \right\rangle \langle \mathbf{z}|}  = \int {d\varphi (\zeta )\left| \zeta  \right\rangle \langle {\zeta}|}  = 1,
\label{unity}
\end{eqnarray}
where the integral measures are defined  by $d\varphi ({\mathbf{z}})
= \prod\limits_k {\frac{{\exp ( - {{\bar z}_k}{z_k}){d^2}{z_k}}}{\pi
},}$ and $d\varphi (\zeta ) = {\exp ( - \bar \zeta \zeta ){d^2}\zeta
}$. As  shown, the bosonic coherent states we use here are not
normalized, and the normalization factors are moved into the
integration measures, which is similar to the Bargmann
representation of the complex space. The application of the coherent
state representation makes the evaluation of path integrals
extremely simple. In the coherent state representation, the
Hamiltonians of the system, the environment, and the interaction
between them are expressed as, respectively
\begin{eqnarray}
\begin{aligned}
{H_S}({\bar{\zeta} },\zeta ) =& \Delta {\bar{\zeta} }\zeta  + \Omega ({\bar{\zeta} } + \zeta ),\\
{H_E}({{\rm{\bar{\mathbf{z}}}} },\mathbf{z}) =& \sum\limits_k {{\Omega _k}\bar{z}_k {z_k}},\\
{H_I}({\bar{\mathbf{z}} },\mathbf{z},{\bar{\zeta} },\zeta ) =& ({g_k}{\bar{\zeta} }{z_k} + g_k^ * \bar{z}_k \zeta ).
\label{representation}
\end{aligned}
\end{eqnarray}
With these notations, we will present  a detailed derivation of the
exact master equation for the reduced density matrix of the system
in the following sections.
\subsection{The influence functional in coherent state representation}
Explicitly, the  density matrix of the whole  system  (the system
plus the environment) obeys the quantum Liouville equation,
$i\partial {\rho _T}(t)/\partial t = [H,{\rho _T}(t)]$, which gives
the formal solution
\begin{eqnarray}
{\rho _T}(t) = \exp ( - iHt){\rho _T}(0)\exp (iHt).
\label{rhoT   }
\end{eqnarray}
In the coherent state representation, by use of Eq.~(\ref{unity}),
${\rho _T}(t)$ can be expressed as
\begin{equation}
\begin{aligned}
&\langle {\zeta _f},{{\bf{z}}_f}|{\rho _T}(t) \left| {{{\zeta '}_f},{{\bf{z}}_f}} \right\rangle\\
=& \int {d\varphi ({\mathbf{z}_i})d\varphi ({\zeta _i})d\varphi ({\mathbf{z}'_i})d\varphi ({\zeta '_i})\left\langle {{{\zeta _f},{\mathbf{z}_f};t}}
 \mathrel{\left | {\vphantom {{{\zeta _f},{\mathbf{z}_f};t} {{\zeta _i},{\mathbf{z}_i};0}}}
 \right. \kern-\nulldelimiterspace}
 {{{\zeta _i},{\mathbf{z}_i};0}} \right\rangle } \\
 &\times \left\langle {{\zeta _i},{\mathbf{z}_i}} \right|{\rho _T}(0)\left| {{\zeta '_i},{\mathbf{z}'_i}} \right\rangle \left\langle {{{\zeta '_i},{\mathbf{z}'_i};0}}
 \mathrel{\left | {\vphantom {{{\zeta '_i},{\mathbf{z}'_i};0} {{\zeta '_i},{\mathbf{z}_i};t}}}
 \right. \kern-\nulldelimiterspace}
 {{{\zeta '_f},\mathbf{{z}}_f;t}} \right\rangle.
 \label{repre}
\end{aligned}
\end{equation}
Assume   the initial density matrix   be factorized into a direct
product of the system and the environment state, i.e.,  ${\rho
_T}(0) = \rho (0) \otimes {\rho _E}(0)$ \cite{Leggett198759}, the
reduced density matrix of the system  is then  given by
\begin{equation}
\begin{aligned}
&\rho (\bar{\zeta} _f ,{\zeta '_f};t) = \int {d\varphi ({\mathbf{z}_f})\left\langle {{\zeta _f},{\mathbf{z}_f}} \right|{\rho _T}(t)\left| {{\zeta '_f},{\mathbf{z}_f}} \right\rangle } \\
& = \int {d\varphi ({\zeta _i})d\varphi ({\zeta '_i})\rho (\bar{\zeta} _i ,{\zeta '_i};0) \cdot J(\bar{\zeta} _f ,{\zeta '_f};t|\bar{\zeta} _i ,{\zeta '_i};0)}.
\label{reducedmatrix}
\end{aligned}
\end{equation}
The next task is to determine the effective propagating function
for the reduced density
matrix \cite{Feynamn196324118,Anastopoulos200062,Ishizaki2008347185},
\begin{equation}
\begin{aligned}
J(\bar{\zeta} _f ,{\zeta '_f};t|\bar{\zeta} _i ,{\zeta '_i};0) =& \int {{D^2}\zeta {D^2}\zeta '\exp \{ i({S_S}[{\bar{\zeta}},\zeta ]} \\
 &- S_S^ * [\bar{\zeta}^\prime ,\zeta '])\} F[{\bar{\mathbf{\zeta}}},\mathbf{\zeta},\bar{\mathbf{\zeta}} ^\prime ,\mathbf{\zeta}'],
\label{JJ}
\end{aligned}
\end{equation}
with ${{S_S}[{\bar{\zeta}},\zeta ]}$ being the action of the system
in the atomic coherent state representation, see Eq.~(\ref{dfg}).
$F[{\bar{\mathbf{\zeta}}},\mathbf{\zeta},\bar{\mathbf{\zeta}}^\prime
,\mathbf{\zeta}']$ is the influence functional which takes  into
account the back-action (in Eq.~(\ref{FJ})) of the environment on
the system.

Assume the environment be initially at zero temperature, i.e., the
initial state of the environment takes,
\begin{eqnarray}
{\rho _E} = {\left| 0 \right\rangle _{BB}}\langle 0|,
\label{rhoE}
\end{eqnarray}
then the influence functional can be solved exactly and we have
\begin{equation}
\begin{aligned}
F[{\bar{\mathbf{\zeta}}},\mathbf{\zeta},\bar{\mathbf{\zeta}}^\prime ,\mathbf{\zeta}'] =& \exp \{ \int_{t_0}^t {d\tau \int_{t_0}^\tau  {d\tau '[f(\tau  - \tau ')} } \\
&(\bar{\zeta}^\prime (\tau ) - {\bar{\zeta}}(\tau ))\zeta (\tau ') + {f^ * }(\tau  - \tau ')\\
&\bar{\zeta}^\prime (\tau ')(\zeta (\tau ) - {\zeta ^\prime }(\tau ))]\},
 \label{finally F}
\end{aligned}
\end{equation}
where
\begin{equation}
\begin{aligned}
f(\tau  - \tau ') =& \sum\limits_k {{{\left| {{g_k}} \right|}^2}{e^{ - i{\Omega _k}(\tau  - \tau ')}}} \\
{\rm{ = }}&\int {d\omega J(\omega ){e^{ - i(\omega  - {\omega _L})(\tau  - \tau ')}}}
\label{ft}
\end{aligned}
\end{equation}
is called the dissipation-fluctuation kernel.

The details of   derivation of Eq.~(\ref{finally F}) can be found in
Appendix.

\section{The exact non-Markovian master equation}

\begin{figure}[h]
\centering
\includegraphics[angle=0,width=0.48\textwidth]{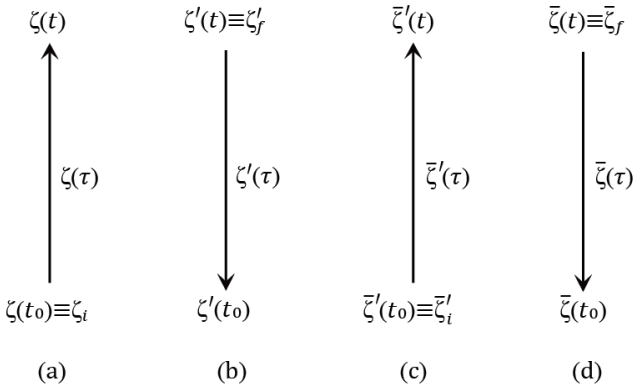}
\caption{(Color online) Schematic illustration  of the four
independent paths denoted by   $\zeta (\tau )$, $\zeta '(\tau )$,
$\bar \zeta '(\tau )$ and $\bar \zeta (\tau )$, respectively.}
\label{bianjietiaojian}
\end{figure}

We now derive the master equation for the reduced density matrix of
the system. Since the effective action after tracing/integrating out
the environmental degrees of freedom, (i.e., combining
Eqs.~(\ref{JJ}) and (\ref{finally F}) together) is in a quadratic
form of the dynamical variables, the path integral (\ref{JJ}) can be
calculated exactly by making use of the stationary path method and
Gaussian integrals \cite{Faddeev1980,Feynman1965}. Substituting
Eq.~(\ref{dfg}) into Eq.~(\ref{JJ}), we have
\begin{equation}
\begin{aligned}
J({{\bar \zeta}_f},{\zeta '_f};t|{\zeta _i},{{\bar \zeta '}_i};0) =& \int {{D^2}\zeta {D^2}\zeta '\exp \{ \frac{1}{2}[{{\bar \zeta }_f}\zeta (t) + \bar \zeta ({t_0}){\zeta _i}} \\
& + \bar \zeta '(t){\zeta'_f} + {{\bar \zeta '}_i}\zeta '({t_0})] - \int_{t_0}^t {d\tau \frac{1}{2}[\bar \zeta \dot \zeta } \\
&- \dot{\bar{\zeta}} \zeta  + \dot{\bar{\zeta}}'\zeta ' - \bar \zeta '\dot \zeta '] + i{H_S}(\bar \zeta ,\zeta )\\
&-i{H_S}(\bar \zeta ',\zeta ')\} F[\bar \zeta ,\zeta ,\bar \zeta ',\zeta '].
\label{FINALLYJ}
\end{aligned}
\end{equation}
To calculate the path integral in Eq.~(\ref{FINALLYJ}),  we use the
stationary phase method \cite{Klauder1979192349,zhang199062867},
which yields  the equations of motion
\begin{equation}
\begin{aligned}
&\dot \zeta (\tau ) + i[\Omega  + \Delta  \cdot \zeta (\tau )] + \int_{t_0}^\tau  {d\tau 'f(\tau  - \tau ')\zeta (\tau ')}  = 0,\\
&\dot \zeta '(\tau ) + i[\Omega  + \Delta  \cdot \zeta '(\tau )] - \int_\tau ^t {d\tau 'f(\tau  - \tau ')\zeta '(\tau ')}\\
& + \int_{t_0}^t {d\tau 'f(\tau  - \tau ')\zeta (\tau ')}  = 0,
\label{fstationary path}
\end{aligned}
\end{equation}
subject to the boundary conditions $\zeta ({t_0}) = {\zeta _i}$ and
$\zeta '(t) = {\zeta '_f}$, respectively. \textbf{ $\bar \zeta
'(\tau )$ and $\bar \zeta (\tau )$ denote the conjugates of $ \zeta
'(\tau )$ and $\zeta (\tau )$, respectively. The equations for these
conjugations can be obtained by     first exchanging $\zeta (\tau )$
and $\zeta '(\tau )$ in Eq.~(\ref{fstationary path}) and taking then
a complex conjugate to these equations. The corresponding boundary
conditions are $\bar \zeta '({t_0}) \equiv \bar \zeta '_i$ and $\bar
\zeta (t) \equiv {\bar \zeta _f}$.  With these   boundary
conditions, we can get the solution of $\zeta (\tau )$ and $\zeta
'(\tau )$. For clarity, we illustrate these notations  in
Fig.~\ref{bianjietiaojian}. Noticing   ${t_0} \le \tau  \le t$, we
 keep in mind that   $\zeta (t)$ in
Fig.~\ref{bianjietiaojian} (a) can be obtained by setting  $\tau  =
t$ and $\zeta '({t_0})$ in Fig.~\ref{bianjietiaojian} (b) can be
obtained by  $\tau  = {t_0}$.  Fig.~\ref{bianjietiaojian} (c) and
(d) is similar, namely,  $\bar \zeta '(t)$ and $\bar \zeta '({t_0})$
can be obtained with  $\tau  = t$ and $\tau  = {t_0}$,
respectively.}

The solution of the integro-differential  Eq.~(\ref{fstationary
path}) can be expressed in terms of two complex functions $u(\tau)$
and $u_1(\tau)$ as
\begin{equation}
\begin{aligned}
\zeta '(\tau )=& {u_1}(\tau )[{\zeta '_f} - \zeta (t)]+\zeta (\tau ) ,\\
\zeta (\tau ) =& u(\tau ){\zeta _i} + h(\tau ),
\label{transformation}
\end{aligned}
\end{equation}
a similar transformation can be written down for their conjugate
variables with the exchange of $\zeta $ with $\zeta'$ for the
boundary values $\bar{\zeta}(t) = \bar{\zeta}_f$ and
$\bar{\zeta}'(t_0)=\bar{\zeta}'_i$. Substituting
Eq.~(\ref{transformation}) into Eq.~(\ref{fstationary path}), we can
obtain the equations of motion for $u(\tau ), {u_1}(\tau )$ and
$h(\tau )$
\begin{equation}
\begin{aligned}
\dot u(\tau ) + i\Delta  \cdot u(\tau ) + \int_{t_0}^\tau  d\tau 'f(\tau  - \tau ')u(\tau ') &= 0, \\
{{\dot u}_1}(\tau ) + i\Delta \cdot{u_1}(\tau ) - \int_\tau ^t d \tau 'f(\tau  - \tau '){u_1}(\tau '){\rm{ }} &= 0, \\
\dot h(\tau ) + i\Delta  \cdot h(\tau ) + \int_{t_0}^\tau  d\tau 'f(\tau  - \tau ')h(\tau ') &=  - i\Omega ,
\label{uu}
\end{aligned}
\end{equation}
subject to the boundary conditions  ${u_1}(t)=1$, $u(t_0)=1$ and
$h(t_0)=0$ with $t_0 \le \tau ,\tau ' \le t.$ By means of Laplace
transform to Eq.~(\ref{uu}), we can easily  find that
\begin{equation}
\begin{aligned}
{u_1}(\tau) = {u^ * }(t - \tau ),h(\tau ) =  - i\Omega \int_{t_0}^\tau  {d\tau 'u(\tau  - \tau ')}.
\label{solution for u}
\end{aligned}
\end{equation}
Now, we set $\tau  = t_0$ in the first equation and $\tau  = t$ in
the second equation of Eq.~(\ref{transformation}), $\zeta (t)$ and
$\zeta '({t_0})$ can be expressed in terms of the boundary
conditions ${\zeta _i}$ and $\zeta'_{f}$
\begin{equation}
\begin{aligned}
\zeta (t) =& u(t){\zeta _i} + h(t),\\
\zeta '(t_0) =& {u^*}(t)[{{\zeta '}_f} - h(t)] + n(t){\zeta _i},
\label{inversezeta}
\end{aligned}
\end{equation}
where $n(t) = 1 - {\left| {u(t)} \right|^2}$. Similarly,
${\bar{\zeta}}(t_0)$ and $\bar{\zeta}'(t)$ can be obtained by
exchanging $\zeta $ and $\zeta'$ in Eq.~(\ref{inversezeta}) and by
taking a complex conjugate to these equations. Finally, substituting
these results with Eq. (\ref{transformation}) into Eq.
(\ref{FINALLYJ}), we obtain the form of the propagating function for
the reduced density matrix
\begin{equation}
\begin{aligned}
J({{\bar \zeta }_f},{\zeta '_f};t|{\zeta _i},{{\bar \zeta '}_i};0)=& \exp \{ u(t)[\bar{\zeta}_{f}  - {h^ * }(t)]{\zeta _i} +{u^ * }(t)\bar{\zeta}'_{i}\\
& \times [\zeta'_{f} - h(t)] +n(t)\bar{\zeta}'_{i}{\zeta _i} + h(t)\bar{\zeta }_f\\
& + {h^ *(t) }\zeta'_{f} - {\left| {h(t)} \right|^2}\}.
\label{finallyJJ}
\end{aligned}
\end{equation}
\textbf{Notice  that the pre-exponential factor in
Eq.~(\ref{finallyJJ})  is one, this is due to the fact that
Eq.~(\ref{finallyJJ}) is the result of integrating out fluctuations
around the stationary path.} Now we can derive the master equation
by computing the time derivative of Eq.~ (\ref{reducedmatrix}).
First, from Eq.~(\ref{finallyJJ}), we can write down the following
identities
\begin{eqnarray}
{\zeta _i}J = \frac{1}{u}\left( {\frac{{\delta J}}{{\overline{\zeta} _f}} - hJ} \right),\bar{\zeta}'_{i}J = \frac{1}{{{u^ * }}}\left( {\frac{{\delta J}}{{{\zeta '_f}}} - {h^ * }J} \right),
\label{zetai}
\end{eqnarray}
which will be used to remove   ${\zeta _i}$ and $\bar{\zeta}'_{i}$
from the time derivative of $J$. After taking time derivative of
Eq.~(\ref{reducedmatrix}) and substituting Eqs.~(\ref{finallyJJ})
and (\ref{zetai}) into it, we obtain the evolution equation
\begin{equation}
\begin{aligned}
\frac{{\partial \rho (\overline{\zeta} _f ,{\zeta '_f})}}{{\partial t}} =& m\bar{\zeta} _f{{\rm P}_1} + {m^ * }{\zeta '_f}{{\rm P}_2} - (m + {m^ * }){{\rm P}_3}\\
& + {m^ * }{h^ * }{{\rm P}_1} + mh{{\rm P}_2} - {{\dot h}^ * }{{\rm P}_1} - \dot h{{\rm P}_2}\\
& - m h\bar{\zeta} _f\rho  - {m^ * }{h^ * }\rho {\zeta '_f} + \dot h\bar{\zeta} _f\rho\\  &+ {{\dot h}^ * }\rho {\zeta '_f},
\label{representationrho}
\end{aligned}
\end{equation}
where,  $m(t) \equiv \frac{{\dot u(t)}}{{u(t)}}$, ${{\rm P}_1}
\equiv \frac{\partial{\rho}}{\partial{\overline\zeta_f}}$, ${{\rm
P}_2} \equiv \frac{{\partial \rho }}{{\partial {\zeta '_f}}}$,
${{\rm P}_3} \equiv \frac{{{\delta ^2}\rho }}{\partial ^2
\overline\zeta _f\zeta_f'}$. By introducing  the following
functional differential relations in the coherent state
representation \cite{Anastopoulos200062,Tu200878235311}
\begin{equation}
\begin{aligned}
&\bar{\zeta} _f{{\rm P}_1} \leftrightarrow {\sigma _ + }
\sigma_-\rho (t),{{\rm P}_2}{\zeta '_f} \leftrightarrow
\rho (t){\sigma _ + }\sigma_-,{{\rm P}_3}
\leftrightarrow \sigma_-\rho (t){\sigma _ + },
\label{234}
\end{aligned}
\end{equation}
we arrive at an exact non-Markovian master equation
\begin{equation}
\begin{aligned}
\frac{{d\rho (t)}}{{dt}} =  - i[H(t),\rho (t)]
+ \gamma (t)[2\sigma_-\rho (t){\sigma _ + }
- \{ {\sigma _ + }\sigma_-,\rho (t)\} ],
\label{finallyrhoo}
\end{aligned}
\end{equation}
with the effective Hamiltonian containing the classical driven field
\begin{eqnarray}
H(t) = s(t){\sigma _ + }\sigma_- + r(t){\sigma _ + } + {r^ * }(t)\sigma_-.
\label{H(t)}
\end{eqnarray}
The renormalized frequency $s(t)$ and the renormalized driving field
$r(t)$ are results of the back-action of the environment. The time
dependent dissipative coefficient $\gamma (t)$ describes the
dissipative non-Markovian  dynamics  due to the interaction between
the system and environment. All these time-dependent coefficients
can be  given explicitly,
\begin{equation}
\begin{aligned}
s(t) =& \frac{i}{2}[m(t) - c.c.],\\
\gamma (t){\rm{ = }}& - \frac{1}{2}[m(t) + c.c.],\\
r(t) =& i[\dot h(t) -h(t)m(t)],
\label{dphiin}
\end{aligned}
\end{equation}
where $u(t)$ and $h(t)$   are determined by  the
integro-differential equations of Eq.~(\ref{uu}). The non-Markovian
effect  is fully manifested   in the integral kernels in
Eq.~(\ref{uu}), which include   the non-local time-correlation
function $f(t)$ of the environment. The non-Markovian memory effect
is   coded into the homogenous non-local time integrals  with the
integral kernel.  In addition, our derivation of the master equation
is fully non-perturbative, which goes beyond the TCL, NZ and
Markovian approximations and includes all effects resulting from the
environment-system couplings.

\section{comparison between the exact and approximate master equations}

\subsection{The Nakajima-Zwanzig and time-convolutionless master equations}

To derive the second-order perturbative master equation, we first go
to the interaction picture, in which the effective Hamiltonian
${H_I}(t)$ in  Eq.~(\ref{H}) can be rewritten as
\begin{eqnarray}
{H_I}(t) = \sigma_-(t){a^\dag }(t) + H.c.,
\label{HI}
\end{eqnarray}
where $\sigma_-(t) = {U^\dag }(t)\sigma_-U(t)$, $U(t) =  {e^{ -
i{H_S}t}}, $ ${a^\dag }(t) = \sum\limits_k {{g_k}a_k^\dag
{e^{i{\Omega _k}t}}} $. The  density operator ${{\bar \rho }_T}(t)$
of the whole system including the system and environment satisfies
the following Liouville equation
\begin{eqnarray}
\dot{\bar{\rho}}_T(t) =  - i[{H_I}(t),{{\bar \rho }_T}(t)].
\label{rhoI}
\end{eqnarray}
Integrating the left and right sides of Eq.~(\ref{rhoI}), we have
\begin{eqnarray}
{{\bar \rho }_T}(t) = {{\bar \rho }_T}(t_0) -  i\int_{t_0}^t
{dt'[{H_I}(t'),{{\bar \rho }_T}(t')]} . \label{rhI}
\end{eqnarray}
Substituting Eq.~(\ref{rhI}) into Eq.~(\ref{rhoI}), we   obtain
\begin{equation}
\begin{aligned}
\dot{\bar{\rho}}_T(t)=&  - i[{H_I}(t),{{\bar \rho }_T}(0)]\\
&- \int_{t_0}^t {dt'[{H_I}(t),[{H_I}(t'),{{\bar \rho }_T}(t')]]} .
\label{rhoI2}
\end{aligned}
\end{equation}
Tracing over the degrees of freedom of the environment, we can
obtain the dynamical equation for the system density matrix $\bar
\rho (t) = T{r_B}{{\bar \rho }_T}(t)$
\begin{equation}
\begin{aligned}
\dot{\bar{\rho}}(t)=&  - iT{r_R}[{H_I}(t),{{\bar\rho}_T}(t_0)]\\
&- T{r_R}\int_{t_0}^t {dt'[{H_I}(t),[{H_I}(t'),{{\bar \rho }_T}(t')]]} .
\label{rhosi}
\end{aligned}
\end{equation}
Let us apply the Born approximation and assume that the reservoir
stays in the vacuum state (\ref{rhoE}) in the dynamics, then we have
\begin{eqnarray}
\dot{\bar{\rho}} (t) =  - T{r_R}\int_{t_0}^t
{dt'[{H_I}(t),[{H_I}(t'),\bar \rho (t') \otimes {\rho _E}]]}.
\label{ohbmnb}
\end{eqnarray}
Notice that
\begin{equation}
\begin{aligned}
\left\langle {a(t){a^\dag }({t_1})} \right\rangle  =& f(t - {t_1}),\\
\left\langle {{a^\dag }(t){a^\dag }({t_1})} \right\rangle
=& \left\langle {a(t)a({t_1})} \right\rangle
= \left\langle {{a^\dag }(t)a({t_1})} \right\rangle  = 0,
\label{dgbbvx}
\end{aligned}
\end{equation}
where $\left\langle A \right\rangle {\rm{ =
T}}{{\rm{r}}_B}\left\langle {A{\rho _E}} \right\rangle  =
\left\langle 0 \right|A\left| 0 \right\rangle_B$, and  substituting
Eq.~(\ref{HI}) into Eq.~(\ref{ohbmnb}), we have
\begin{equation}
\begin{aligned}
\dot{\bar{\rho}}(t) = \int_{t_0}^t {dt'f(t - t')[} {\rm{ }}
\sigma_- (t')\bar \rho (t'),{\sigma _+}(t)] + H.c..
\label{NZ2}
\end{aligned}
\end{equation}
By transforming Eq.~(\ref{NZ2}) back into  the Schr\"{o}dinger
picture, we   obtain
\begin{equation}
\begin{aligned}
{{\dot \rho }_{NZ}} =&  - i[{H_S},{\rho _{NZ}}(t)]
+ \int_{{t_0}}^t {dt'\{ f(t - t')} [U(t - t')\\
& \times {\sigma _ - }{\rho _{NZ}}(t'){U^\dag }(t - t'),{\sigma _ +
}] + H.c.\}. \label{RHOnz2}
\end{aligned}
\end{equation}
The non-Markovian master equation (\ref{RHOnz2}) is in the standard
form of the Nakajima-Zwanzig (NZ) equation $\dot \rho (t) = \int_0^t
{dt'f(t,t')\rho (t')}$ \cite{Nakajima195820,Zwanzig196033}, where
the NZ kernel $f(t,t')$ is of the time-translationally-invariant
form $f(t-t')$.

Note that Eq.~(\ref{ohbmnb}) is in a form of delayed
integro-differential equation and thus it is a time-nonlocal master
equation. It is worth reminding  that the other systematically
perturbative non-Markovian master equation that is local in time can
be derived from the time-convolutionless projection operator
formalism \cite{Breuer2002,Breuer1999591633,Yan1998582721}. Now, we
go to the details. Under a similar assumption, i.e., the factorized
initial system-reservoir density matrix, the second-order
time-convolutionless master equation in the interaction picture can
be obtained
\cite{Breuer2002,Breuer1999591633,Yan1998582721,Ferraro200980042112,Xu20011143868,
Schroder2006124084903,Liu200776022312,Haikka201081052103,Haikka2010014047}
\begin{equation}
\begin{aligned}
\dot{\bar{\rho}}(t)=  - T{r_R}\int_{t_0}^t {dt'[{H_I}(t),[{H_I}(t'),
\bar \rho (t) \otimes {\rho _E}]]}.
\label{tcl2}
\end{aligned}
\end{equation}
Substituting Eq.~(\ref{HI}) into  Eq.~(\ref{tcl2}) and using
Eq.~(\ref{dgbbvx}), we   transform  Eq.~(\ref{NZ2}) back into the
Schr\"{o}dinger picture and obtain,
\begin{equation}
\begin{aligned}
{{\dot \rho }_{TCL}} =&  - i[{H_S},{\rho _{TCL}}(t)] + \int_{t_0}^t {dt'\{ f(t - t')} \\
&\times[\sigma_- (t' - t){\rho _{TCL}}(t){\sigma_+} - {\sigma_+}\\
&\times\sigma_- (t' - t){\rho _{TCL}}(t)]+ H.c.\}.
\label{tcll22}
\end{aligned}
\end{equation}
We note here that obtaining the time-convolutionless non-Markovian
master equation perturbatively up to  second order in the coupling
by the use of  the time-convolutionless projection operator
technique is equivalent to obtaining it by replacing $\bar \rho
(t')$ with $\bar \rho (t)$ in Eq.~(\ref{ohbmnb})
\cite{Breuer2002,Breuer1999591633,
Yan1998582721,Ferraro200980042112,Xu20011143868,
Schroder2006124084903,Liu200776022312,Haikka201081052103,Haikka2010014047}.
One may wonder if the second order time-nonlocal master equation
(\ref{RHOnz2}) is more accurate than the second-order
time-convolutionless master equation (\ref{tcll22}). In the
following, using the exact master equation, we   show that the TCL
approach (\ref{tcll22}) works better than the NZ one (\ref{RHOnz2})
for a wide range of parameters.
\subsection{Comparison to the Nakajima-Zwanzig and time-convolutionless master equations}
We now analyze the characteristics of the damped  driven two-level
systems, by comparing the exact dynamics with that from the NZ and
TCL master equations. Our purpose is to shed light on the
performances of two master equations and to point out their ranges
of validity. As stressed in the introduction, without the exact
master equation, it is difficult to examine the  range of validity
for these master equations.

We assume that the system couples to a reservoir with detuning
and the reservoir has a Lorentzian spectral
density \cite{Li201081062124,Breuer2002,Shen201388033835,Haikka201081052103}
\begin{eqnarray}
J(\omega ) = \frac{\Gamma }{{2\pi }}\frac{{{\lambda ^2}}}{{{{(\omega  - {\omega _0} + \delta )}^2} + {\lambda ^2}}},
\label{spectral density}
\end{eqnarray}
where $\delta  = {\omega _0} - {\omega _c}$ is the  detuning of
$\omega _c$ to $\omega _0$, and $\omega _c$ is the center frequency
of the cavity. It is worth noting that the parameter $\lambda $
defines the spectral width of the reservoir and is connected to the
reservoir correlation time ${\tau _R} = {\lambda ^{ - 1}}$. The
parameter $\Gamma $ can be shown to be related to the decay of the
system in the Markovian limit  with a flat spectrum. The relaxation
time scale is ${\tau _L} = {\Gamma ^{ - 1}}$.

\textbf{The Markovian dynamics usually describes  a situation
where the coupling strength between the system and the environment
is very weak}, and the characteristic correlation time ${\tau _R}$
of the environment is sufficiently shorter than that of the system
${\tau _L}$, i.e.,
\begin{eqnarray}
{\tau _R} \ll {\tau _L},
\label{markovlimit}
\end{eqnarray}
equivalently, the spectrum of the reservoir takes $J(\omega ) =
\frac{\Gamma }{{2\pi }}$, which leads to a  Markovian dynamics. The
reservoir has no memory effect on the evolution of the system. Then
according to Eq.~(\ref{ft}), we have
\begin{eqnarray}
f(t) = \Gamma \delta (t).
\label{markovainft}
\end{eqnarray}
Substituting Eq.~(\ref{markovainft}) into the first equation of
Eq.~(\ref{uu}), we reduce the solution of $u(t)$   to
\begin{eqnarray}
u(t) = {e^{ - i\Delta t - \frac{\Gamma }{2}t}},
\label{uut}
\end{eqnarray}
i.e., all the coefficients in  Eq.~( \ref{dphiin}) are constants,
\begin{eqnarray}
s(t) = \Delta ,r(t) = \Omega ,\gamma (t) = \Gamma.
\label{reducessrt}
\end{eqnarray}
The exact master equation (\ref{finallyrhoo}) is then  reduced to
Markovian master equation
\cite{Breuer2002,Gardiner2000,Shatokhin2000174157}
\begin{equation}
\begin{aligned}
\frac{{d\rho (t)}}{{dt}} =&  - i[\Delta {\sigma_+}\sigma_-  + \Omega {\sigma _x},\rho (t)] + \frac{\Gamma }{2}[2\sigma_- \rho (t){\sigma _ + }\\
 &- {\rm{\{ }}{\sigma _ + }\sigma_- ,\rho (t)\} ],
\label{Markovianmaster}
\end{aligned}
\end{equation}
where the decoherence rates are time independent. This gives the
standard Lindblad form for the Markovian dynamics. When
\begin{eqnarray}
{\tau _R} \ge {\tau _L}
\label{nonmarkovlimit}
\end{eqnarray}
is satisfied, the strong non-Markovian effect plays an important
role and the dynamics must be described by the exact master equation
(\ref{finallyrhoo}).

Now we calculate the two-time correlation functions $f(t - t')$ by
substituting Eq.~(\ref{spectral density}) into Eq.~(\ref{ft})
\begin{eqnarray}
f(t - t') = \frac{1}{2}\lambda \Gamma \exp [ - (\lambda  + i\Delta  - i\delta )(t - t')].
\label{lorenzft}
\end{eqnarray}
It is clear that the bandwidth $\lambda$ is inversely proportional
to  the memory time of reservoir. For this correlation function $f(t
- t')$, Eq.~(\ref{uu}) can be easily solved by use of
Eq.~(\ref{lorenzft}), the solution reads,
\begin{eqnarray}
u(t) = k(t) \times \left[ {\cosh \left( {\frac{{dt}}{2}}  \right) +
\frac{{\lambda  - i\delta }}{d}\sinh \left( {\frac{{dt}}{2}}
\right)} \right],
\end{eqnarray}
where $k(t) = {e^{ - (\lambda  + 2i\Delta  - i\delta )t/2}}$  and $d
= \sqrt {{{(\lambda  - i\delta )}^2} - 2\Gamma \lambda } $.

In order to calculate $U(t)$ and $\sigma_-(t)$  in
Eqs.~(\ref{RHOnz2}) and (\ref{tcll22}), we calculate  the
eigenstates of the free system Hamiltonian ${H_S}$,
\begin{equation}
\begin{aligned}
\left| {{\phi _{\lambda 1}}} \right\rangle  = \frac{1}{{\sqrt 2 }}(\sqrt {1 + \sin \theta } \left| e \right\rangle  + \sqrt {1 - \sin \theta } \left| g \right\rangle ),\\
\left| {{\phi _{\lambda 2}}} \right\rangle  = \frac{1}{{\sqrt 2 }}(\sqrt {1 - \sin \theta } \left| e \right\rangle  - \sqrt {1 + \sin \theta } \left| g \right\rangle ),
\label{twobasic}
\end{aligned}
\end{equation}
the corresponding eigenvalues are    ${\lambda _1} = (\Delta  +
{W_0})/2$ and ${\lambda _2} = (\Delta  - {W_0})/2$. Here ${W_0} =
\sqrt {{\Delta ^2} + 4{\Omega ^2}} $, $\theta = acr\tan (\Delta
/2\Omega )$. Straightforward algebra yields,
\begin{equation}
\begin{aligned}
\sigma_-(t) =& {e^{i{H_S}t}}\sigma_- {e^{ - i{H_S}t}}  = \sum\limits_{j,k = 1}^2 {{\sigma _{jk}}{e^{it({\lambda _j} - {\lambda _k})}}\left| {{\phi _{\lambda j}}} \right\rangle \left\langle {{\phi _{\lambda k}}} \right|} ,\\
U(t) =& \sum\limits_{j = 1}^2 {{e^{i{\lambda _j}t}}\left| {{\phi _{\lambda j}}} \right\rangle \left\langle {{\phi _{\lambda j}}} \right|} ,
\label{Usigma}
\end{aligned}
\end{equation}
where ${\sigma _{jk}} = \left\langle {{\phi _{\lambda j}}}
\right|\sigma_- \left| {{\phi _{\lambda k}}} \right\rangle $. Now
let us concentrate on the average  $\left\langle {{\sigma _z}}
\right\rangle $, i.e., on the probability difference of finding the
system in the atomic excited and ground levels. To examine the
validity of the two approximate approaches we explore three
different regimes by changing the width $\lambda$ of the Lorentzian
spectral density. This investigation will allow us to estimate in
which cases the non-Markovian master equations are efficient in the
description of the system  dynamics.
\begin{figure}[h]
\centering
\includegraphics[angle=0,width=0.48\textwidth]{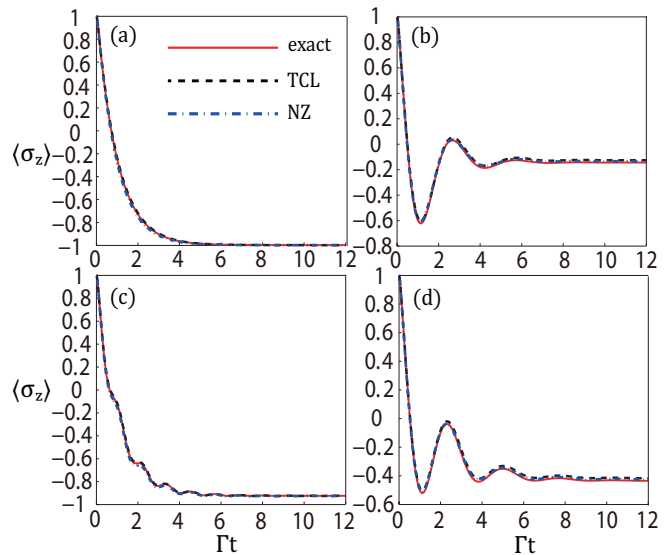}
\caption{(Color online) The time evolution of  the population
difference $\left\langle {{\sigma _z}} \right\rangle $ for the
system initially   in the excited state $\left| e \right\rangle$
versus the dimensionless parameter $\Gamma t$. The red line,
black-dashed line, and blue-dashed-dotted line denote the exact
  Eq.~(\ref{finallyrhoo}), TCL Eq.~(\ref{tcll22}), and NZ
Eq.~(\ref{RHOnz2}) master quations, respectively. The width of the
Lorentzian spectral density is $\lambda  = 25\Gamma$. The other
parameters chosen are $\Delta  = 0.3\Gamma,\Omega  = 0.02\Gamma,\delta  = 0.01\Gamma$
for (a), $\Delta  = 0.3\Gamma,\Omega  = \Gamma,\delta  = 0.01\Gamma$ for (b), $\Delta
= 5\Gamma,\Omega  = \Gamma,\delta  = 0.01\Gamma$ for (c), $\Delta  = \Gamma,\Omega  =
\Gamma,\delta  = 10\Gamma$ for (d).} \label{nonmar:}
\end{figure}
Fig.~\ref{nonmar:} shows a comparison among the  exact, TCL, and NZ
master equations with  large bandwidth $\lambda  = 25\Gamma .$ We
find that the results given by the TCL (\ref{tcll22}) and NZ
(\ref{RHOnz2}) are in good agreement with those obtained by the
exact master equation (\ref{finallyrhoo}) for any time scales. In
this case, both TCL and NZ give a very good description for the
dynamics. They indeed provide us with the same results, which are
very close  to the Markovian dynamics; see the discussion in
Eq.~(\ref{Markovianmaster}). In addition, in such cases the TCL
master equation which is easier to solve might be preferred to use
because it is a  time-local first order differential equations.
\begin{figure}[h]
\centering
\includegraphics[angle=0,width=0.48\textwidth]{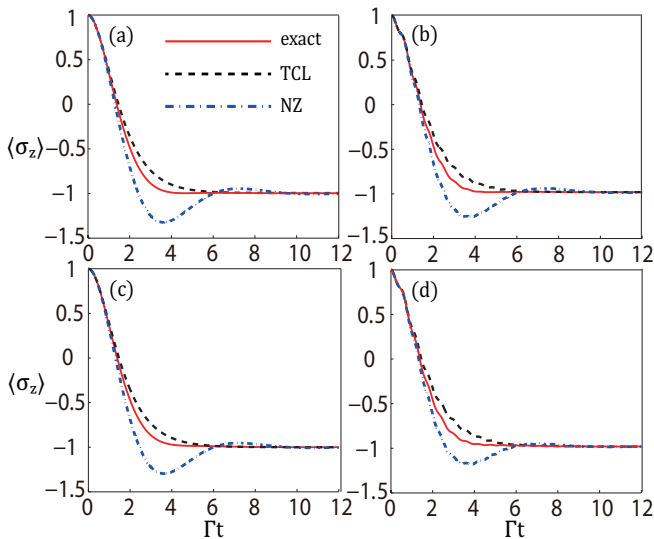}
\caption{(Color online) $\left\langle {{\sigma _z}} \right\rangle $
versus the dimensionless parameter $\Gamma t$. The width of the
Lorentzian spectrum  is $\lambda  =\Gamma$. The results are obtained
by the exact (red line), TCL (black-dashed line), and
NZ (blue-dashed-dotted line) solutions. The other parameters chosen
are $\Delta  = 0.3\Gamma,\Omega  = 0.02\Gamma,\delta  = 0.01\Gamma$ for (a), $\Delta =
10\Gamma,\Omega  = \Gamma,\delta  = 0.01\Gamma$ for (b), $\Delta  = 10\Gamma,\Omega  =
0.02\Gamma,\delta  = 0.2\Gamma$ for (c), $\Delta  = 10\Gamma,\Omega  = \Gamma,\delta  =
0.2\Gamma$ for (d).}\label{litnonmar:}
\end{figure}

We set the same quantity $\lambda  = \Gamma$ in
Fig.~\ref{litnonmar:}. Clearly, the results given by the TCL
(\ref{tcll22}) and NZ (\ref{RHOnz2}) are in good agreement
with those obtained by the exact expression (\ref{finallyrhoo})
in a short-time scale, but they deviate  from each other in a
long-time scale. Especially considering the long-time behavior, the
NZ equation leads to a non-physical result. For times longer than
some critical values, the solution for the population difference
$\left\langle {{\sigma _z}} \right\rangle $ cannot represent a
physical result, because the absolute value of $\left\langle
{{\sigma _z}} \right\rangle $ is larger than $1$. We therefore can
conclude that for this range of parameters the TCL equation gives a
better description of the dynamics because it reflects all the
qualitative characteristics of the exact expression.

\begin{figure}
\centering
\includegraphics[scale=0.36]{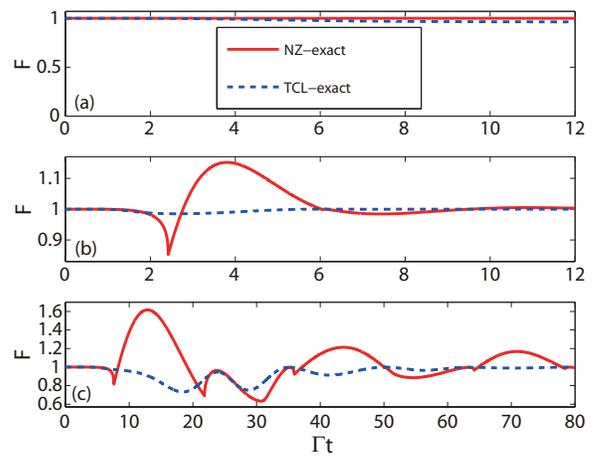}
\caption{(Color online) Comparison of  the density matrices  obtained by solving
the TCL and NZ master equation with the one by exact master
equation. We quantify the difference by the fidelity defined by
$F(\rho_1,\rho_2)=Tr \sqrt{\rho_1^{\frac 1 2}\rho_2\rho_1^{\frac 1 2
}}$. The results show that the density matrix given by TCL is always
better than that given by NZ master equation. The parameters in
(a),(b) and (c) are chosen  as the same as in Fig.\ref{nonmar:}-(a),
Fig.\ref{litnonmar:}-(a) and Fig.\ref{strnonmar:}-(a), respectively.
}\label{fig3:}
\end{figure}
One may wonder if this observation depends on the quantity plotted.
To clarify this point, we plot the fidelity of the density matrix
from the exact master equation to these from TCL and NZ master
equations in Fig.~\ref{fig3:}. The results suggest that the TCL
master equation is indeed better than the NZ for a wide range of
parameters.

\begin{figure}[h]
\centering
\includegraphics[angle=0,width=0.483\textwidth]{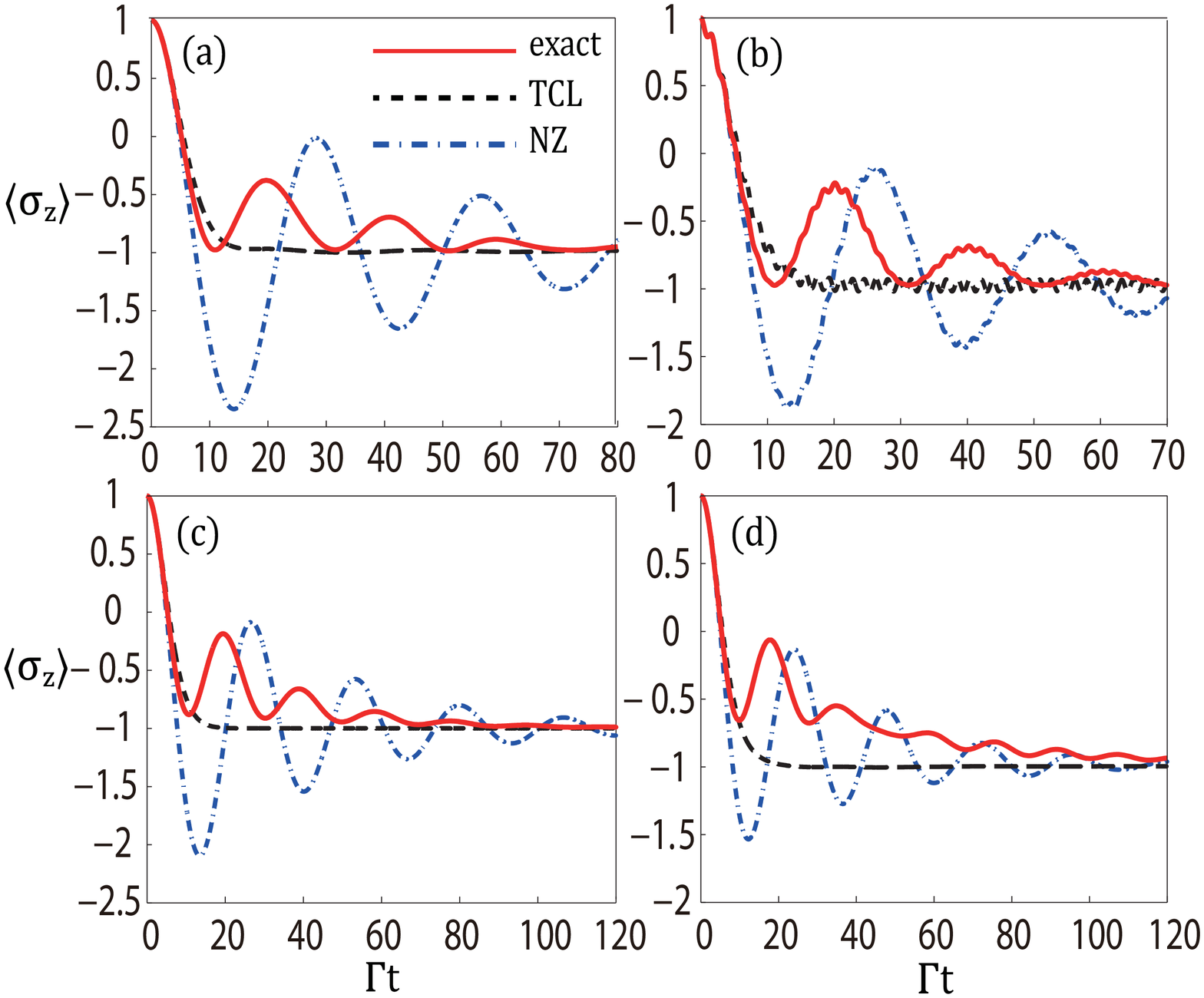}
\caption{(Color online) $\left\langle {{\sigma _z}} \right\rangle $
versus time $\Gamma t$. The results are obtained by the exact (red
line), TCL (black-dashed line), and NZ (blue-dashed-dotted line)
solutions. The parameters chosen are $\lambda  =0.05\Gamma$, $\Delta
= 0.3\Gamma,\Omega  = 0.02\Gamma,\delta  = 0.01\Gamma$ for (a), $\Delta  = 3.5\Gamma,\Omega
= 0.4\Gamma,\delta  = 0.01\Gamma$ for (b), $\Delta  = 10\Gamma,\Omega  = 0.02\Gamma,\delta
= 0.08\Gamma$ for (c), $\Delta  = 0.3\Gamma,\Omega  = 0.02\Gamma,\delta  = 0.14\Gamma$ for
(d).}\label{strnonmar:}
\end{figure}

In Fig.~\ref{strnonmar:}, we choose the parameter  $\lambda  =
0.05\Gamma $, which, according to Eq.~(\ref{lorenzft}), corresponds
to very strong reservoir correlations and very long memory effect.
We find again that a good agreement among all the three approaches
in the short-time scale, but in this case the TCL approximation
works not so good. The dynamics of the TCL master equation
(black-dashed line) does not succeed to follow the oscillations
given by the exact expression (red line). The NZ approach has the
same problem that it can not conserve the positivity of the
density matrix (i.e., the absolute value of $\left\langle {{\sigma
_z}} \right\rangle $ exceeds $1$). Thus in this case two approximate
methods are not suitable to describe the dynamics of the driven
two-level system.

\textbf{Before closing this section,  we present a discussion on the
function  $f\left( {\tau  - \tau '} \right)$ in Eq.~(\ref{ft}).
Concretely,  we examine mathematically the validity to extend the
lower limit of the integration  from $0$ to $ - \infty $. We will
explore three different regimes characterized by the width $\lambda
$ in the  spectral density in the following.}

\begin{figure}[h]
\centering
\includegraphics[angle=0,width=0.489\textwidth]{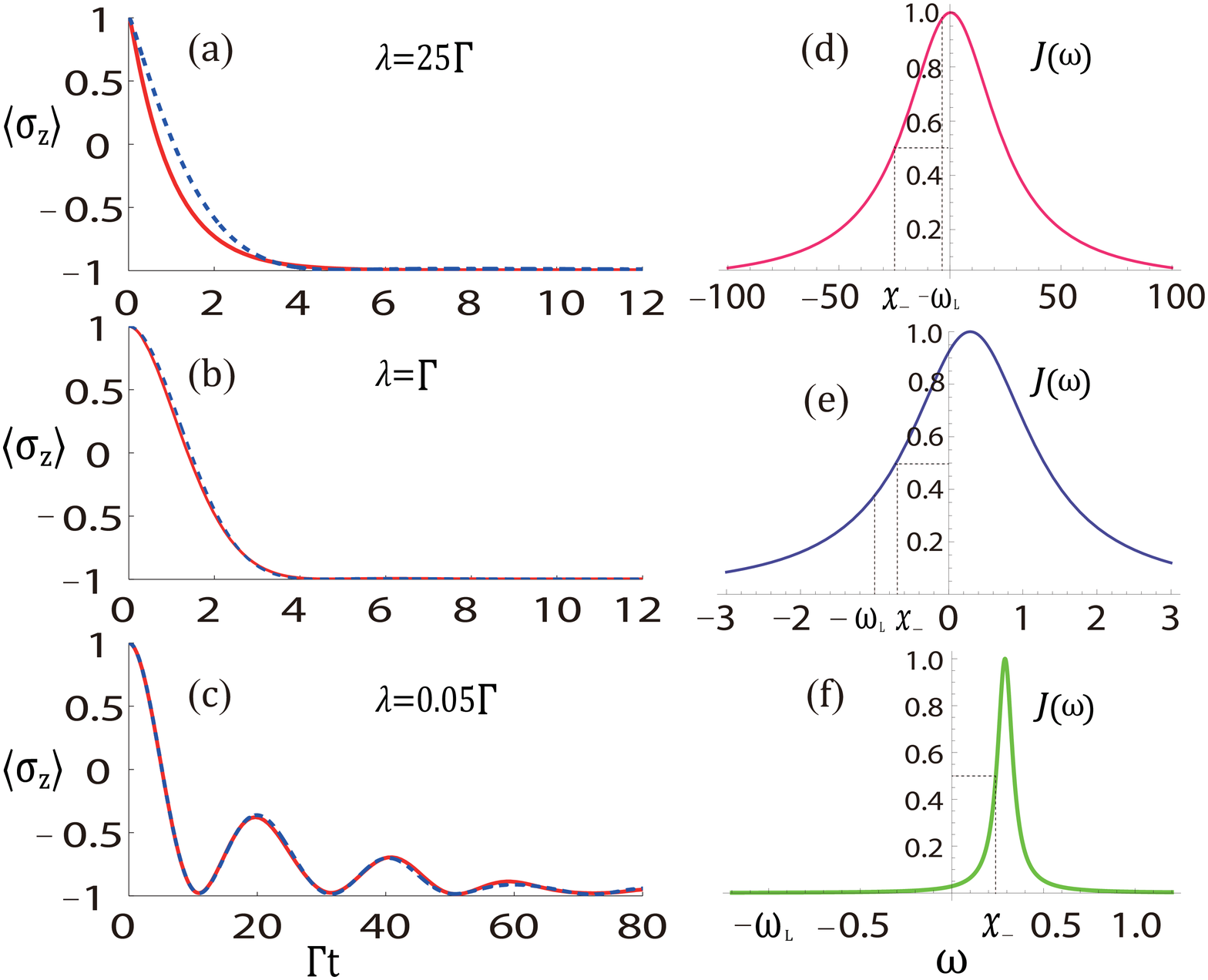}
\caption{(Color online)\textbf{ $\left\langle {{\sigma _z}}
\right\rangle $ given by  the exact master equation
(\ref{finallyrhoo}) as a function of time. The purpose of this
figure is to show the difference in $\left\langle {{\sigma _z}}
\right\rangle$ caused by different lower limits of the integral of
kernel (\ref{fttt}). The red and blue-dashed lines correspond to
lower limits  $-\omega_L$ and $-\infty$, respectively. The
Lorentzian spectral density $J(\omega )$ (in units of
$\Gamma/{2\pi}$) in (d), (e) and (f) correspond respectively to
results shown in (a), (b) and (c). $x_-$ denotes the left location
of the half-height of the spectral density. The parameters in (a),
(b) and (c) are chosen as the same as in Fig.\ref{nonmar:}-(a),
Fig.\ref{litnonmar:}-(a) and Fig.\ref{strnonmar:}-(a), respectively.
Notice that $\Delta=\omega_0-\omega_L=0.3\Gamma$ in Eq.~(\ref{H}),
we set $\omega_0=1.3\Gamma$ and
$\omega_L=\Gamma$.}}\label{zero-infinity:}
\end{figure}

\textbf{In Fig.~\ref{zero-infinity:},  we show a comparison between
results with two different lower limits  in  the integration
(\ref{ft}) with the   spectral density given in Eq. (\ref{spectral
density}), the simulation is performed  for the exact dynamics
described  by Eq.~(\ref{finallyrhoo}). Fig.~\ref{zero-infinity:} (a)
is for the integration with   lower limit     $- \infty$, which is
slightly different from that with lower limit 0. In
Fig.~\ref{zero-infinity:} (b) and (c), the results with  lower limit
 $- \infty$  are in good agreement with that obtained with lower
limit  0.}

\textbf{This numerical result can be explained as follows.  When  we
change $\omega \to \omega  - {\omega _L}$, Eq.~(\ref{ft}) becomes
}\begin{eqnarray}
\begin{aligned}
f\left( {\tau  - \tau '} \right) =
\int_{ - {\omega _L}}^\infty  {d\omega J\left( \omega  \right)}
{e^{ - i\omega (\tau  - \tau ')}}
\label{fttt}
\end{aligned}
\end{eqnarray}
\textbf{with}
\begin{eqnarray}
J\left( \omega  \right) = \frac{\Gamma }{{2\pi }}
\frac{{{\lambda ^2}}}{{{{\left( {\omega  - \Delta  + \delta } \right)}^2} + {\lambda ^2}}},
\label{jwww}
\end{eqnarray}
\textbf{this tells us that the frequency ${{\omega _L}}$  affects
only the lower limit  of the integral (\ref{fttt}) when $\Delta$ is
fixed. Define ${x_ \pm }= \Delta - \delta  \pm \lambda $
representing the position of half-height of the Lorentzian spectral
density (\ref{jwww}), we thought that the integral of $J(\omega)$
over $\omega$ from $-\infty$ to $\infty$ can be approximately
replaced by the same integral but from $x_-$ to $x_+$. With this
approximation, we find that ${x_ - } = -24.71\Gamma$ and $\lambda =
25\Gamma$ in Fig.~\ref{zero-infinity:} (a). Clearly, $x_ -$  is much
smaller than $-{{\omega _L}}$, thus the integral of $J(\omega)$ over
$\omega$ from $x_-$ to $-{{\omega _L}}$ can not be ignored [see
Fig.~\ref{zero-infinity:} (d)]. This explains the difference  of the
two curves in Fig.~\ref{zero-infinity:} (a). On the contrary,
$\lambda  = \Gamma$, ${x_ - } = -0.71\Gamma$ in
Fig.~\ref{zero-infinity:} (b),  and  $\lambda  = 0.05\Gamma$, ${x_ -
} = 0.24\Gamma$ in Fig.~\ref{zero-infinity:} (c). $x_-$ is larger
than $-{{\omega _L}}$ in both cases of (b) and (c). Thus, the
integral from $-{{\omega _L}}$ to $x_-$ can be ignored [see
Fig.~\ref{zero-infinity:} (e) and (f)]. As a result,  the two lines
in both (b) and (c) are in good agreement.}

\textbf{The above discuss suggests that it is  reasonable to extend
the lower limit  of the integral of Lorentzian spectral $J\left(
\omega \right)$ from $0$ to $ - \infty $.}

\section{validity of secular approximation in time-convolutionless master equations}

Taking advantage of the exact expression for the dissipative
dynamics of the open driven two-level system, we have shown that the
TCL approach can reveal all the characteristics of the
non-Markovian dynamics for a range of parameters much wider than the
results that the NZ equation gives, this is  physically reasonable,
since the latter may violate the positivity condition on the density
matrix for the reservoir correlations which are not very strong.
Therefore through comparing with the exact non-Markovian master
equation (\ref{finallyrhoo}), we can investigate the validity of the
secular approximation based on time-convolutionless master equation
(\ref{tcll22}).

We now use the orthonormalized basis (\ref{twobasic}) and these
relations (\ref{Usigma}) to derive explicitly the
time-convolutionless master equations (\ref{tcll22})  as follows
\begin{eqnarray}
\dot \rho  =  - i[{H_S} - H_1,\rho ] + D(\rho ) + {D_1}(\rho ),
\label{tcl2expend}
\end{eqnarray}
with
\begin{equation}
\begin{aligned}
H_1 =g_0^2{Q_0}(t)S_z^2 +g_2^2{Q_{ + 1}}(t)S_-{S_+}+g_1^2{Q_{ - 1}}(t){S_+}S_-,
\label{Hprimee}
\end{aligned}
\end{equation}
  which describes a small shift
in the energy   of the two-level system. The above new operators are
defined as $S_-{\rm{ = }}\left| {{\phi _{\lambda 2}}} \right\rangle
\left\langle {{\phi _{\lambda 1}}} \right|$, ${S_ + }{\rm{ =
}}\left| {{\phi _{\lambda 1}}} \right\rangle \langle {\phi _{\lambda
2}}|,$ and ${S_z}{\rm{ = }}\left| {{\phi _{\lambda 1}}}
\right\rangle \left\langle {{\phi _{\lambda 1}}} \right| - \left|
{{\phi _{\lambda 2}}} \right\rangle \left\langle {{\phi _{\lambda
2}}} \right|$, then the dissipative superoperator $D(\rho)$ in
Eq.~(\ref{tcl2expend}) can be written in a Lindblad form
\begin{equation}
\begin{aligned}
D(\rho ) =& g_1^2{P_{ - 1}}(t)[2S_- \rho {S_+} - \{ {S_+}S_-,\rho \} ]\\
 &+ g_2^2{P_{ + 1}}(t)[2{S_+}\rho S_- - \{ S_-{S_+},\rho \} ]\\
 &+ g_0^2{P_{ 0}}(t)[2{S_z}\rho {S_z} - \{ S_z^2,\rho \} ],
\label{domig}
\end{aligned}
\end{equation}
where the coefficients ${g_0} = \Omega /{W_0}$,  ${g_1} = ({W_0} +
\Delta )/(2{W_0})$, ${g_2} = ({W_0} - \Delta )/(2{W_0})$, ${W_0} =
\sqrt {{\Delta ^2} + 4{\Omega ^2}} $. The second dissipator
$D_1(\rho)$ in Eq.~(\ref{tcl2expend}) has a more complicated form
and contains the contribution of the so-called nonsecular terms,
\begin{equation}
\begin{aligned}
{D_1}(\rho ) =&{g_0}{R_0}(t)[{g_2}({S_z}\rho S_- - S_-{S_z}\rho ) + {g_1}({S_ + }\\
&\times{S_z}\rho- {S_z}\rho{S_ +})]+ {g_2}{R_1}(t)[{g_0}({S_ + }\rho {S_z}\\
&- {S_z}{S_ + }\rho )- {g_1}{S_ + }\rho {S_ + }]+ {g_1}{R_{ - 1}}(t)[{g_0}\\
&\times({S_z}S_-\rho  - S_-\rho {S_z})- {g_2} S_- \rho S_-]+ H.c..
\label{nonsecular}
\end{aligned}
\end{equation}
For TCL master equations, the non-Markovian  effects are contained
in the time-dependent coefficients ${P_m}(t),{Q_m}(t),$ and
${R_m}(t)$, with $m \in \{ {\rm{ + ,0,}} - \}$. The time-dependent
coefficient reads
\begin{equation}
\begin{aligned}
{R_m}(t) =& \int_0^t {dt'\int {d\omega J(\omega )\exp [i({M_m} - \omega )(t-t')]} },
\label{coefficients}
\end{aligned}
\end{equation}
where ${M_m} = {\omega _L} - m{W_0}$. The other  coefficients take
${P_m}(t) = {\rm{Re[}}{R_m}(t)]$ and ${Q_m}(t) =  -
{\rm{Im[}}{R_m}(t)].$ Conventionally, the nonsecular terms included
in the dissipator ${D_1}(\rho )$ are neglected in the secular
approximation. In order to investigate the effects of the nonsecular
terms on the non-Markovian dynamics, we focus on two regimes
identified by the mutual relationship between the system
characteristic time   and the reservoir correlation time.

The time-dependent coefficient (\ref{coefficients}) for the driven
two-level system in a Lorentzian reservoir can be calculated
explicitly using Eq.~(\ref{lorenzft})
\begin{eqnarray}
{R_m}(t) = \frac{{\Gamma \lambda }}{{\lambda  +  i{N_m}}}\left\{ {1
- \exp [ - (\lambda  + i{N_m})t]} \right\}, \label{rmt}
\end{eqnarray}
with
\begin{eqnarray}
{N_m}{\rm{ = }}\Delta {\rm{ - }}\delta {\rm{ + }}m{W_0}.
\label{nm}
\end{eqnarray}
We can see from Eq.~(\ref{rmt}) that when  $\min \left[ {\left| {{N_
+ }} \right|,\left| {{N_0}} \right|,\left| {{N_ - }} \right|}
\right] \gg \lambda,$ namely, the relaxation time ${\tau _R} =
{\lambda ^{ - 1}}$ of the reservoir correlation is very large
compared to the typical timescale defined as ${\tau _S} = {[\min
(\left| {{N_ + }} \right|,\left| {{N_0}} \right|,\left| {{N_ - }}
\right|)]^{ - 1}}$, i.e.
\begin{eqnarray}
{\tau _R} \gg {\tau _S}
\label{tairtaus}
\end{eqnarray}
is satisfied, oscillating terms (\ref{nonsecular}) (that containing
${R_m}(t)$) may be neglected as $t$ increases, since rapid
oscillations have no contribution to the dynamics  on the timescale
of the relaxation, this constitutes the secular approximation.

When
\begin{eqnarray}
{\tau _R} \le {\tau _S},
\label{nonsecularcondition}
\end{eqnarray}
we cannot neglect the nonsecular terms (\ref{nonsecular}) in the
master equation (\ref{tcl2expend}) in the dynamics of the driven
two-level system. Therefore in this case, we can no longer obtain a
simple expression for the system. The master equation of the system
is no longer in the time-dependent Lindblad form.

Examining Eqs.~(\ref{markovlimit}) and (\ref{tairtaus}), we can
summarize the comparison of the nonsecular with the secular
approximation in the following Table \ref{table:}, which shows the
validity regimes for secular and nonsecular approximation in TCL,
Markovian and non-Markovian regimes, respectively.
\begin{table}[h]
\centering \caption{Comparison of regimes of secular  and nonsecular
approximation in TCL for Markovian and non-Markovian regimes,
respectively.}
\includegraphics[scale=0.426]{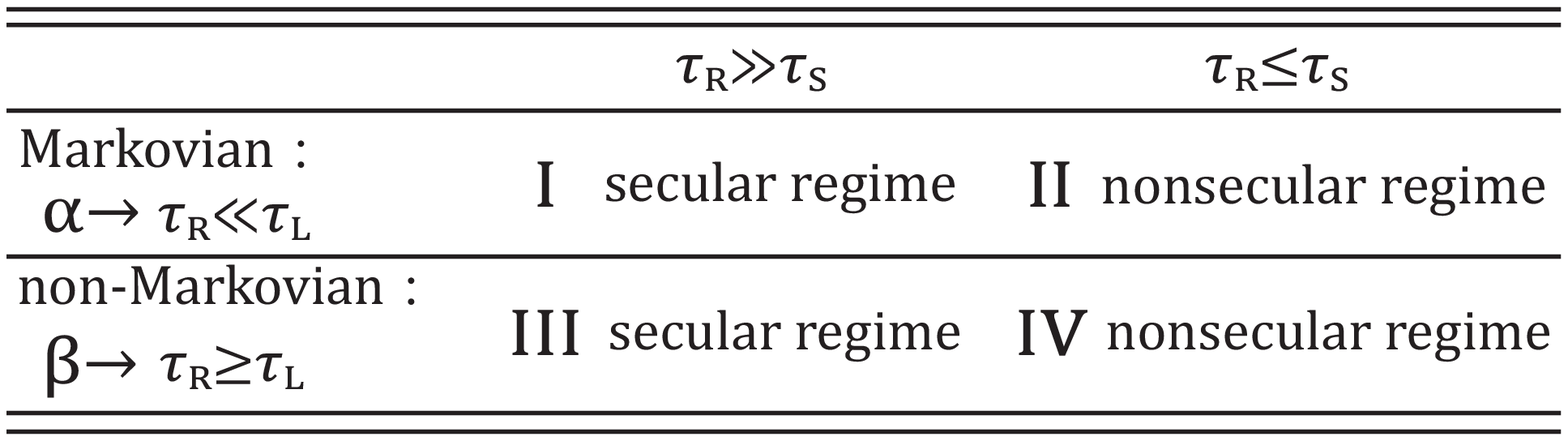}
\label{table:}
\end{table}
From Table \ref{table:}, we can divide the  time dependent dynamics
into two regimes, labeled by $\alpha$ and $\beta$, i.e., Markovian
and non-Markovian regimes, respectively.
\begin{figure}[h]
\centering
\includegraphics[scale=0.391]{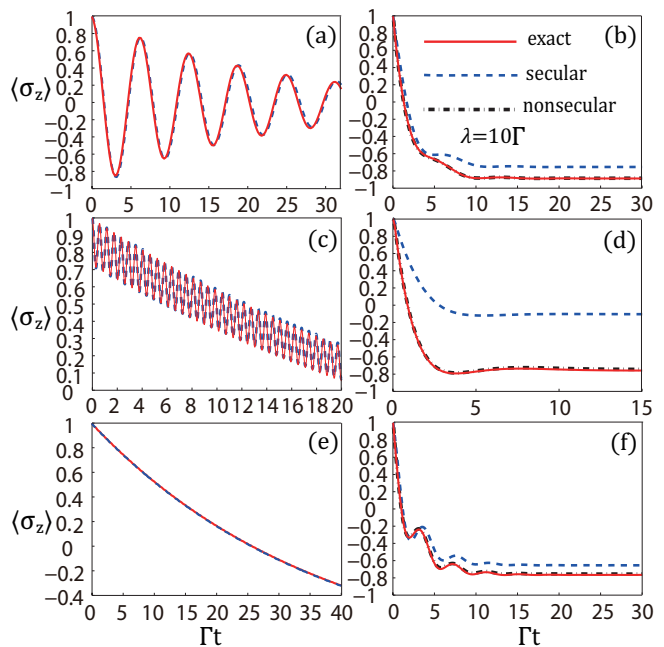}
\caption{(Color online) This plot shows the  comparison of the
secular approximation (regime $\rm I$) [(a), (c), and (e)] and  nonsecular terms (regime $\rm II$) [(b),
(d), and (f)] in Markovian regime $\alpha$ in Table \ref{table:}.
The red line , blue-dashed line, and black dashed-dotted line denote
the exact expression Eq.~(\ref{finallyrhoo}), the secular
approximation Eq.~(\ref{tcl2expend}) neglecting the nonsecular terms
(\ref{nonsecular}), and the nonsecular Eq.~(\ref{tcl2expend})
containing (\ref{nonsecular}), respectively. Parameters chosen are
$\lambda=10\Gamma, \Delta  = 0,\Omega  = 0.5\Gamma,\delta  = 40\Gamma$ for (a),
$\Delta  = 0.5\Gamma,\Omega  = 0.2\Gamma,\delta  = 10\Gamma$ for (b), $\Delta  =
10\Gamma,\Omega  = 2\Gamma,\delta  = 60\Gamma$ for (c), $\Delta  = 0.1\Gamma,\Omega  =
0.2\Gamma,\delta  = 5\Gamma$ for (d), $\Delta  = 10\Gamma,\Omega  = 0.2\Gamma,\delta  = 60\Gamma$
for (e), $\Delta  = \Gamma,\Omega  = 0.5\Gamma,\delta  = 10\Gamma$ for (f).}
\label{Marksecu:}
\end{figure}
In regime $\alpha$, i.e., Markovian regime, we can see that the
results given by the regime $\rm I$ under the secular approximation
in the TCL Eq.~(\ref{tcl2expend}) are in good agreement with those
obtained by the exact master equation Eq.~(\ref{finallyrhoo}) when
the weak coupling condition (\ref{markovlimit}) and the secular
approximation (\ref{tairtaus}) are simultaneously satisfied [see
Figs.~\ref{Marksecu:} (a), \ref{Marksecu:} (c), and \ref{Marksecu:}
(e)]. When the parameters simultaneously satisfy
Eqs.~(\ref{markovlimit}) and (\ref{nonsecularcondition}) [see
Figs.~\ref{Marksecu:} (b), \ref{Marksecu:} (d), and \ref{Marksecu:}
(f)], i.e., the regime (\rm II), the dynamics of the TCL master
equation (\ref{tcl2expend}) involving the nonsecular terms
Eq.~(\ref{nonsecular}) are in good agreement with those obtained by
the exact expression (\ref{finallyrhoo}), but the results obtained
by the secular approximation have serious deviations from those
obtained by the exact solution Eq.~(\ref{finallyrhoo}). This
difference comes from the nonsecular terms (\ref{nonsecular}), which
are ignored in the regime (\rm II).

\begin{figure}[h]
\centering
\includegraphics[scale=0.3904]{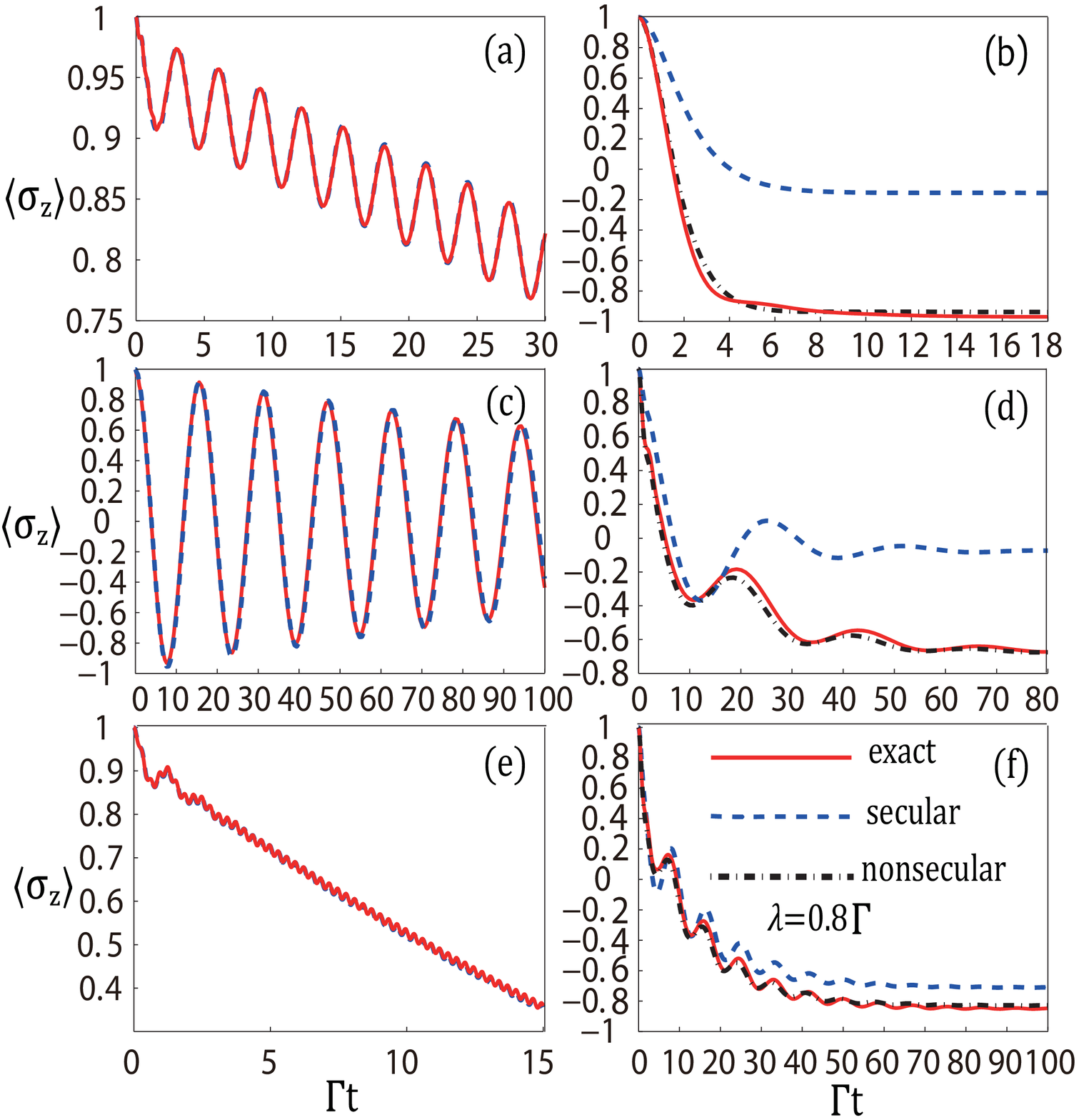}
\caption{(Color online) This plot shows the   comparison of the
secular approximation (regime $\rm III$) [(a), (c), and (e)] with
nonsecular terms (regime $\rm IV$) [(b), (d), and (f)] in
non-Markovian regime $\beta$ in Table \ref{table:}. The red line,
blue-dashed line, and black dashed-dotted line denote the exact
master equation Eq.~(\ref{finallyrhoo}), the secular approximation
Eq.~(\ref{tcl2expend}) neglecting the nonsecular terms
(\ref{nonsecular}), and the nonsecular Eq.~(\ref{tcl2expend})
containing (\ref{nonsecular}), respectively. Parameters chosen are
$\lambda=0.8\Gamma, \Delta  = 2\Gamma,\Omega  = 0.2\Gamma,\delta  = 15\Gamma$ for (a),
$\Delta  = 0.04\Gamma,\Omega  = 0.06\Gamma,\delta  = 0.4\Gamma$ for (b), $\Delta  =
0,\Omega  = 0.2\Gamma,\delta  = 10\Gamma$ for (c), $\Delta  = 0.05\Gamma,\Omega  =
0.1\Gamma,\delta  = 1.8\Gamma$ for (d), $\Delta  = 20\Gamma,\Omega  = \Gamma,\delta  = 5\Gamma$
for (e), $\Delta  = 0.5\Gamma,\Omega  = 0.2\Gamma,\delta  = 2.5\Gamma$ for (f).}
\label{nonmarsecu:}
\end{figure}

Examining the non-Markovian regime labeled by  $\beta$ in Table
\ref{table:}, we find that the results given by the secular
approximation Eq.~(\ref{domig}) in the regime $\rm III$   are in
good agreement with those obtained by the exact expression
Eq.~(\ref{finallyrhoo}) when the strong coupling condition
(\ref{nonmarkovlimit}) and the secular approximation
(\ref{tairtaus}) are simultaneously satisfied [see
Figs.~\ref{nonmarsecu:} (a), \ref{nonmarsecu:} (c), and
\ref{nonmarsecu:} (e)]. When the parameters satisfy simultaneously
Eqs.~(\ref{nonmarkovlimit}) and (\ref{nonsecularcondition}) [see
Figs.~\ref{nonmarsecu:} (b), \ref{nonmarsecu:} (d), and
\ref{nonmarsecu:} (f)], i.e., in the regime \rm IV, the dynamics of
the TCL master equation (\ref{tcl2expend}) involving the nonsecular
terms Eq.~(\ref{nonsecular}) are in good agreement with those
obtained by the exact one (\ref{finallyrhoo}). However, the results
obtained by the secular approximation have serious deviations from
the exact solution Eq.~(\ref{finallyrhoo}). The same observation can
be found in the regime $\rm II$.

From Figs.~\ref{Marksecu:} and \ref{nonmarsecu:}, we can learn that
the non-Markovian effect occurs when   $\lambda $ is small. The
non-Markovian regime $\beta$ transits to the Markovian regime
$\alpha$ when   $\lambda$ is large. Therefore by manipulating
$\lambda$ we can control the crossover from non-Markovian to
Markovian processes and vice versa. This provides us with a method
to manipulate the non-Markovian dynamics in the driven two-level
system.

Now we turn to discuss   the  positivity and complete positivity of
the reduced dynamics given by the TCL master equation. The
non-Markovian  TCL master equation  derived  in this paper is not of
the Lindblad form, even in the secular regime discussed in Sec.V,
therefore, both the  positivity and the complete positivity of the
reduced  dynamics  can not be guaranteed. In other words, the
Lindblad-Gorini-Kossakowski-Sudarshan
theorem\cite{Lindblad481191976,Gorini178211976} that ensures the
positivity can not be satisfied in general, indicating that the
dynamics given by the TCL master equation might not be physical for
all range of  parameters.

Nevertheless, the parameters chosen (in fact, it is wide range of
parameters) in this paper assure the positivity of the reduced
dynamics given by the TCL master equations. This can be understood
as follows. For the driven qubit in the TCL approximation, the
necessary and sufficient condition for complete positivity and
positivity is given by (for details, see
Ref.\cite{Haikka201081052103})
\begin{eqnarray}
2\alpha (t) + \beta (t) \ge 0, \label{cpp}
\end{eqnarray}
where
\begin{equation}
\begin{aligned}
\alpha (t) =& 2\int_0^t {d\tau [g_1^2{P_{ - 1}}(\tau ) + g_2^2{P_{ + 1}}(\tau ) + 4g_0^2{P_0}(\tau )]} ,\\
\beta (t) =& 4\int_0^t {d\tau [g_1^2{P_{ - 1}}(\tau ) + g_2^2{P_{ + 1}}(\tau )]} .
\label{Hprimgee}
\end{aligned}
\end{equation}
Now  back to the Sec. IV,  we stress that the necessary and
sufficient condition (\ref{cpp}) for complete positivity is
satisfied for the parameters chosen  in Fig.~(\ref{litnonmar:}) and
(\ref{strnonmar:}) (not for a very long time). Therefore, for a wide
range of parameters, the complete positivity of the reduced dynamics
is guaranteed. Hence our conclusion, i.e., the TCL equation gives a
better description of the dynamics, holds true  for a wide range of
parameters.  It is important to remind that theoretical descriptions
of non-Markovian open quantum systems are often based on a series of
assumptions and approximations without which it would not be
possible to tackle the problem of the description of the dynamics in
simple analytic terms. But those approximations plague almost all
approximated reduced dynamics and lead them to break the complete
positivity required for reduced dynamics. Therefore the observation
here is available for short times and certain ranges of parameters.

\section{The case with non-Lorentzian spectrum}

\begin{figure}[h]
\centering
\includegraphics[scale=0.42]{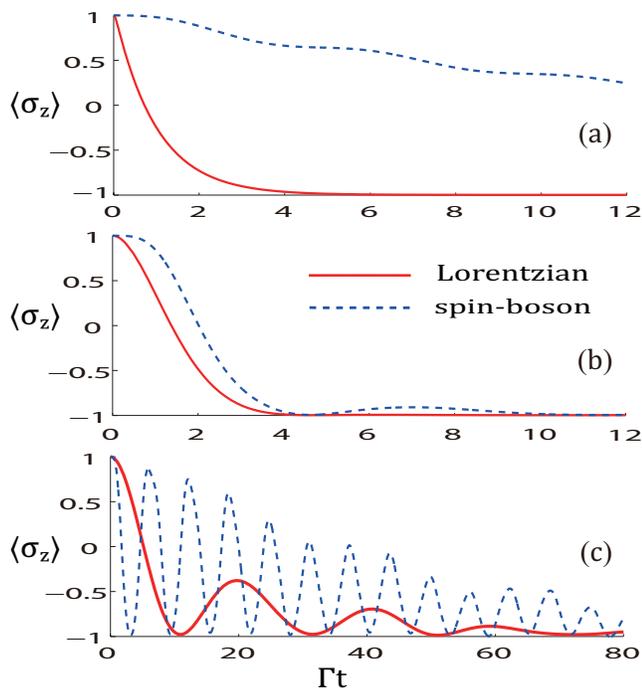}
\caption{(Color online) This plot shows the comparison of the exact dynamics (\ref{finallyrhoo}) for Lorentzian (red-line) and spin-boson(blue-dashed line) spectral density. The parameters in
(a), (b) and (c) are chosen  as the same as in Fig.\ref{nonmar:}-(a),
Fig.\ref{litnonmar:}-(a) and Fig.\ref{strnonmar:}-(a), respectively. The other parameter chosen is $M=5\Gamma$.}
\label{spin-boson:}
\end{figure}

\textbf{Note that the spectral density  ${J_{SB}}(\omega ) $ is
proportional to the imaginary part of the dynamical susceptibility
$\tilde \chi (\omega )$ of a damped harmonic oscillator, in this
section, we present a numerical simulation  for $\left\langle\sigma_z(t)\right\rangle$
adapting a different spectral density, e.g.,  spin-boson spectral
density \cite{Weiss2008,Thoss20011152991}, }\begin{eqnarray}
{J_{SB}}(\omega ) = \frac{1}{M}\frac{{\omega \lambda }}{{{{({\omega
^2} - \omega _0^2)}^2} + {\omega ^2}{\lambda ^2}}}. \label{JSB}
\end{eqnarray}
\textbf{ In Fig.~(\ref{spin-boson:}), we plot the time evolution of
the population difference $\left\langle {{\sigma _z}} \right\rangle$
for three typical spectral width $\lambda$. Interestingly, in
Fig.~\ref{spin-boson:} (a), i.e., for large $\lambda=25\Gamma$ the
population difference  $\left\langle {{\sigma _z}} \right\rangle $
decays monotonically for both spin-boson  and Lorentzian spectral
density, the difference is that the former decay more slowly than the
latter. This corresponds to  the Markovian case, see the discussion
in Eq.~(\ref{Markovianmaster}). For $\lambda=\Gamma$, small
oscillations can be observed  in the case with spin-boson spectral
density, while it is not   obvious in the case with Lorentzian
spectral density [see Fig.~\ref{spin-boson:} (b)]. For small
$\lambda=0.05\Gamma$, oscillations in the population difference can
be found  in both cases with  spin-boson spectral density and
Lorentzian spectral density[see Fig.~\ref{spin-boson:} (c)]. These
oscillations correspond to a rapid exchange of energy and
information between the two-level atom and    reservoir.}

\textbf{Spectral density is a key feature for environments. It
characterizes the correlation among the particles in the environment
and determines the dynamics of open system, as we show in this
section.}

\section{Conclusion}
For a driven two-level  quantum system, secular and weak coupling
approximations break down when the system-environment coupling
varies significantly on the scale of the Rabi frequency. In this
paper,  we avoid these approximations and have studied the
non-Markovian dynamics of the driven two-level system coupled to a
bosonic reservoir at zero temperature. Making use of the
Feynman-Vernon influence functional theory in the coherent state
representation, we derive an exact non-Markovian master equation for
the driven two-level system. We compare this exact master equation
with the other equations describing non-Markovian dynamics, i.e.,
the Nakajima-Zwanzig and the time-convolutionless non-Markovian
master equation, it is found that the TCL approach is valid for a
range of parameters much wider than the NZ master equation. This is
reasonable  since the latter may violate the positivity of dynamical
map when the correlation in the reservoir is strong. By using the
exact master equation, we also have given the analytical condition
of validity of the secular approximation and show how it depends on
the environmental spectral density, we found that the nonsecular
terms have significant corrections to results obtained by the
secular approximation when the relaxation  time of the environment
is less than or equal to that of the system, i.e. ${\tau _R} \le
{\tau _S}$.

The limitation of this representation is the state of the bath, here
we only consider the bath initially at vacuum. Although the zero
temperature case is problematic for getting reduced dynamics as the
bath correlation functions may decay slowly, the zero-temperature
reservoir is a good approximation for many problems in physics. For
the reservoir initially at thermal states, the question becomes
complicated, since the influence functional in the Feynman-Vernon
influence functional theory is very involved.

\section*{ACKNOWLEDGMENTS}
This work is supported by the NSF of China under Grants No.
11175032.

\appendix*
\section{DERIVATION OF THE INFLUENCE FUNCTIONAL}

The propagating function controlling the time evolution  of the
reduced density matrix is given by Eq.~(\ref{JJ}), where the
generalized Feynman-Vernon influence functional is defined by
\begin{equation}
\begin{aligned}
F[{\bar{\zeta}},\zeta,{\bar{\zeta}}^\prime ,\zeta']
{\rm{ = }}&\int {d\varphi ({\mathbf{z}_f})d\varphi
({\mathbf{z}_i})d\varphi ({\mathbf{z}'_i})} {D^2}\mathbf{z}{D^2}\mathbf{z}'\\
 &\times {\rho _E}(\bar{\mathbf{z}}_i ,{\mathbf{z}'_i};0)\exp
 \{ i({S_E}[{\bar{\mathbf{z}}},\mathbf{z}]\\
&- S_E^ * [\bar{\mathbf{z}}',\mathbf{z}'] + {S_I}
[{\bar{\mathbf{z}}},\mathbf{z},{\bar{\zeta} },\zeta ]\\
& -{S_I^*}[\bar{\mathbf{z}}',\mathbf{z}',\bar{\zeta}',\zeta'])\},
\label{FJ}
\end{aligned}
\end{equation}
where ${S_S}, {S_I}$ and ${S_E}$ are the actions  corresponding to
${H_S}, {H_I}$ and ${H_E}$, respectively,
\begin{equation}
\begin{aligned}
{S_S}[{\bar{\zeta}},\zeta ] =&  - i[\bar{\zeta} _f\zeta (t)
+ {\bar{\zeta}}(t_0){\zeta _i}]/2 + \int_{t_0}^t {d\tau \{ }
i[{\bar{\zeta}}(\tau )\\
& \times \dot \zeta (\tau ) - {\dot{\bar{\zeta}} }(\tau )
\zeta (\tau )]/2 - {H_S}({\bar{\zeta} },\zeta )\},\\
{S_E}[{\bar{\mathbf{z}}},\mathbf{z}] =& \sum\limits_k { - i}
\bar{z}_k{z_k}(t) + \int_{t_0}^t {d\tau [i\bar{z}_k{{\dot z}_k}(\tau)} \\
& - {H_E}({\bar{\mathbf{z}}},\mathbf{z})],\\
{S_I}[{\bar{\mathbf{z}}},\mathbf{z},{\bar{\zeta}},\zeta ]
=&  - \int_{t_0}^t {d\tau {H_I}[{\bar{\mathbf{z}}},
\mathbf{z},{\bar{\zeta}},\zeta ]} .\label{dfg}
\end{aligned}
\end{equation}
All the functional integrations are worked  out  over paths
${\bar{\mathbf{z}}}(\tau ),\mathbf{z}(\tau ),{\bar{\zeta}}(\tau )$,
and $\zeta (\tau )$, the endpoints are  ${\bar{\mathbf{z}}}(t)
\equiv \bar{\mathbf{z}}_f,\mathbf{z}(t_0) \equiv
{\mathbf{z}_i},{\bar{\zeta}}(t) \equiv {\zeta _f}$, and $\zeta (t_0)
\equiv {\zeta _i}.$

Now we can calculate explicitly the influence  functional of our
model using the coherent state path-integral formalism. Substituting
Eq.~(\ref{representation}) into the actions of Eq.~(\ref{dfg}), we
obtain the explicit form of the propagator. The path integral of the
environmental part in the propagator can be exactly done by the stationary
phase method \cite{Klauder1979192349,zhang199062867} with the
boundary conditions ${z_k}(t_0) = {z_{ki}}$ and $\bar{z}_k(t) =
\bar{z}_{kf}$. This method needs  the equations of motion of the
path,
\begin{equation}
\begin{aligned}
{{\dot z}_k}+i{\Omega _k}{z_k}=-ig_k^ * \zeta,
{{\dot{\bar{z}}}}_k-i{\Omega_k}\bar{z}_k=i{g_k}{\bar{\zeta}},
\label{station phase}
\end{aligned}
\end{equation}
where $\zeta $ and ${\bar{\zeta}}$ are treated as external sources.
By formally integrating Eq.~(\ref{station phase}), we obtain
(\ref{station phase})
\begin{equation}
\begin{aligned}
{z_k}(\tau ) =& {z_{ki}}{e^{ - i{\Omega _k}\tau }} - ig_k^ * \int_0^\tau  {d\tau '{e^{ - i{\Omega _k}(\tau  - \tau ')}}\zeta (\tau ')} ,\\
\bar{z}_k(\tau ) =& \bar{z}_{kf}{e^{i{\Omega _k}(\tau  - t)}} + i{g_k}\int_\tau ^t {d\tau '{e^{i{\Omega _k}(\tau  - \tau ')}}\bar{\zeta} (\tau ')}.
\label{integratezk}
\end{aligned}
\end{equation}
By taking  the reservoir to  be initially at zero temperature
(\ref{rhoE}), i.e., ${\rho _E}(\bar{\mathbf{z}}_i
,{\mathbf{z}'_i};0)=1$, we finally  can obtain Eq.~(\ref{finally F})
after  substituting the result and Eq.~(\ref{integratezk}) into
Eq.~(\ref{FJ}).


\begin{references}
\bibitem{Alicki2007717} R. Alicki and K. Lendi, \textit{Quantum Dynamical Semigroups and Applications}, Lecture Notes in physics, Vol. \textbf{717}, 2nd ed. (Springer, Berlin, 2007)

\bibitem{Breuer2002} H.-P. Breuer and F. Petruccione, \textit{The Theory of Open Quantum Systems} (Oxford University Press, Oxford, UK, 2002).

\bibitem{Weiss2008} U. Weiss, \textit{Quantum Dissipative Systems}, 3rd ed. (World Scientific Press, Singapore, 2008)

\bibitem{DiVincenzo1998393} D. P. DiVincenzo, Nature \textbf{393}, 113 (1998).

\bibitem{Knill2001409} E. Knill, R. Laflamme, and G. J. Milburn, Nature \textbf{409}, 46 (2001).

\bibitem{Cirac199959} J. I. Cirac, A. K. Ekert, S. F. Huelga, and C. Macchiavello, Phys. Rev. A \textbf{59}, 4249 (1999).

\bibitem{DiVincenzo200048} D. P. DiVincenzo, Fortschr. Phys. \textbf{48}, 771 (2000).

\bibitem{Cirac199778} J. I. Cirac, P. Zoller, H. J. Kimble, and H. Mabuchi, Phys. Rev. Lett. \textbf{78}, 3221 (1997).

\bibitem{Duan200367} L.-M. Duan, A. Kuzmich, and H. J. Kimble, Phys, Rev, A \textbf{67}, 032305 (2003).

\bibitem{Gardiner2000} C. W. Gardiner and P. Zoller, \textit{Quantum Noise} (Springer-Verlag, Berlin, Germany, 2000).

\bibitem{Scully1997} M. O. Scully and M. S. Zubairy, \textit{Quantum Optics} (Cambridge University Press, Cambridge, UK, 1997).

\bibitem{Walls1994} D. F. Walls and G. J. Milburn, \textit{Quantum Optics} (Springer-Verlag, Berlin, 1994).

\bibitem{Carmichael1993} H. J. Carmichael, \textit{An Open Systems Approach to Quantum Optics}, Lecture Notes in Physics m18 (Springer-Verlag, Berlin, 1993).

\bibitem{Mandel1995} L. Mandel and E. Wolf, \textit{Optical Coherence and Quantum Optics}
 (Cambridge University Press, England, 1995).

\bibitem{Weissbluth1988} M. Weissbluth, \textit{Photon-Atom Interactions} (Academic Press, Boston, 1989).

\bibitem{Vogel1994} W. Vogel and D. G. Welsch, \textit{Lectures on Quantum Optics}
(Akademie Verlag, Berlin, 1994).

\bibitem{Compagno1995} G. Compagno, R. Passante, and F. Persico, \textit{Atom-Field Interactions and Dressed Atom} (Cambridge University Press, Cambridge, 1995).

\bibitem{Caldeiral1983149} A. O. Caldeira and A. J. Leggett, Ann. Phys. (N.Y.) \textbf{149}, 374 (1983).

\bibitem{Leggett198759} A. J. Leggett, S. Chakravarty, A. T. Dorsey, M. P. A. Fisher, A. Garg, W. Zwerger, Rev. Mod. Phys. \textbf{59}, 1 (1987).

\bibitem{Majer200594} J. B. Majer, F. G. Paauw, A. C. J. ter Haar, C. J. P. M. Harmans,
and J. E. Mooij, Phys. Rev. Lett. \textbf{94}, 090501 (2005).

\bibitem{Berkley2003300} A. J. Berkley, H. Xu, R. C. Ramos, M. A. Gubrud, F. W.
Strauch, P. R. Johnson, J. R. Anderson, A. J. Dragt, C. J. Lobb,
and F. C. Wellstood, Science \textbf{300}, 1548 (2003).

\bibitem{Pashkin2003421} Y. A. Pashkin, T. Yamamoto, O. Astafiev, Y. Nakamura, D. V. Averin, and J. S. Tsai, Nature (London) 421, 823 (2003).

\bibitem{Bellomo200799} B. Bellomo, R. Lo Franco, and G. Compagno, Phys. Rev. Lett. \textbf{99}, 160502 (2007).

\bibitem{Nielsen2000}M. A. Nielsen and I. L. Chuang, \textit{Quantum Computation and
Quantum Information} (Cambridge University Press, Cambridge, UK, 2000).

\bibitem{Barenco199574} A. Barenco, D. Deutsch, A. Ekert, and R. Jozsa, Phys. Rev.
Lett. \textbf{74}, 4083 (1995).

\bibitem{Biercuk2009458} M. J. Biercuk, H. Uys, A. P. VanDevender, N. Shiga, W. M. Itano, and J. J. Bollinger, Nature (London) \textbf{458}, 996 (2009).

\bibitem{Das200942} S. Das and G S Agarwal, J. Phys. B \textbf{42} 205502 (2009).

\bibitem{Sinaysky200878} I. Sinaysky, F. Petruccione, and D. Burgarth, Phys. Rev. A \textbf{78}, 062301 (2008).

\bibitem{Anastopoulos200062} C. Anastopoulos and B. L. Hu, Phys. Rev. A \textbf{62}, 033821 (2000).
\bibitem{Feynamn196324118} R. P. Feynman and F. L. Vernon, Ann. Phys. (N.Y.) \textbf{24}, 118 (1963).

\bibitem{Caldeira1983121587} A. O. Caldeira and A. J. Leggett, Physica A \textbf{121}, 587 (1983).

\bibitem{Hu1992452843} B. L. Hu, J. P. Paz, and Y. Zhang, Phys. Rev. D \textbf{45}, 2843
(1992).

\bibitem{zhang199062867} W. M. Zhang, D. H. Feng, and R. Gilmore, Rev. Mod. Phys.
\textbf{62}, 867 (1990).

\bibitem{Klauder1979192349} J. R. Klauder, Phys. Rev. D \textbf{19}, 2349 (1979).


\bibitem{Karrlein199755153} R. Karrlein and H. Grabert, Phys. Rev. E \textbf{55}, 153 (1997).

\bibitem{Haake198532} F. Haake and R. Reibold, Phys. Rev. A \textbf{32}, 2462 (1985).

\bibitem{An20093241737} J.-H. An, Y. Yeo, and C. H. Oh, Ann. Phys. (NY) \textbf{324}, 1737
     (2009).

\bibitem{An200776042127} J. H. An and W. M. Zhang, Phys. Rev. A \textbf{76}, 042127 (2007).

\bibitem{An200990317} J. H. An, M. Feng, and W. M. Zhang, Quantum. Inf. Comput. \textbf{9}, 0317 (2009).

\bibitem{Lucke199911110843} A. Lucke, C. H. Mak, and J. T. Stockburger£¬ J. Chem. Phys. \textbf{111}, 10843 (1999).

\bibitem{Chow200877011112} C.-H. Chou, T. Yu, and B. L. Hu, Phys. Rev. E \textbf{77}, 011112 (2008).

\bibitem{Paz2008100220401} J. P. Paz and A. J. Roncaglia, Phys. Rev. Lett. \textbf{100}, 220401 (2008).

\bibitem{Paz200979032102} J. P. Paz and A. J. Roncaglia, Phys. Rev. A 79, 032102 (2009).

\bibitem{Tu200878235311} M. W. Y. Tu and W. M. Zhang, Phys. Rev. B \textbf{78}, 235311 (2008).

\bibitem{Tu20098631} M. W. Y. Tu, M. T. Lee, and W. M. Zhang, Quant. Info. Proc. \textbf{8}, 631 (2009).

\bibitem{Jin201012083013} J. S. Jin, M. T.W. Tu,W. M. Zhang, and Y. J. Yan, New J. Phys.
\textbf{12}, 083013 (2010).

\bibitem{Tan201183} H. T. Tan and W. M. Zhang, Phys. Rev. A \textbf{83}, 032102 (2011).

\bibitem{Lei20123271408} C. U. Lei and W. M. Zhang, Ann. Phys. \textbf{327}, 1408 (2012).

\bibitem{zhang2012109170402} W.-M. Zhang, P.-Y. Lo, H.-N. Xiong, M.W.-Y. Tu, and F. Nori, Phys. Rev. Lett. \textbf{109}, 170402 (2012).

\bibitem{Chaturvedi197935} S. Chaturvedi and F. Shibata, Z. Phys. B \textbf{35}, 297 (1979).

\bibitem{Shibata197717} R. Shibata, Y. Takahashi and N. Hashitsume, J. Stat. Phys. \textbf{17} 171 (1977).

\bibitem{Prataviera201387} G. A. Prataviera, A. C. Yoshida, and S. S. Mizrahi, Phys. Rev. A \textbf{87}, 043831 (2013).

\bibitem{Nakajima195820} S. Nakajima, Prog. Theor. Phys. \textbf{20}, 948 (1958).

\bibitem{Zwanzig196033} R. Zwanzig, J. Chem. Phys. \textbf{33}, 1338 (1960).

\bibitem{zhang201387032117} J. Zhang, Y.-X. Liu, R.-B. Wu, K. Jacobs, and F. Nori,
     Phys. Rev. A \textbf{87}, 032117 (2013).

\bibitem{Breuer1999591633} H.-P. Breuer, B. Kappler, and F. Petruccione, Phys. Rev. A \textbf{59}, 1633 (1999).

\bibitem{Yan1998582721} Y. J. Yan, Phys. Rev. A \textbf{58}, 2721 (1998).

\bibitem{Ferraro200980042112} E. Ferraro, M. Scala, R. Migliore, and A. Napoli, Phys. Rev. A \textbf{80}, 042112 (2009).

\bibitem{Xu20011143868} R. X. Xu and Y. J. Yan, J. Chem. Phys. \textbf{114}, 3868 (2001).

\bibitem{Schroder2006124084903} M. Schr\"{o}er, U. Kleinekath\"{o}fer, and M. Schreiber, J. Chem. Phys. \textbf{124}, 084903 (2006).

\bibitem{Liu200776022312} K.-L. Liu and H.-S. Goan, Phys. Rev. A \textbf{76}, 022312 (2007).

\bibitem{Haikka201081052103} P. Haikka and S. Maniscalco, Phys. Rev. A \textbf{81}, 052103 (2010).

\bibitem{Haikka2010014047} P. Haikka, Phys. Scr. \textbf{2010}, 014047 (2010).

\bibitem{Cahill1999591538} K. E. Cahill and R. J. Glauber, Phys. Rev. A \textbf{59}, 1538 (1999).

\bibitem{Glauber1963131} R. J. Glauber, Phys. Rev. \textbf{131}, 2766 (1963).

\bibitem{Shresta200571022109} S. Shresta, C. Anastopoulos, A. Dragulescu, and B. L. Hu, Phys. Rev. A \textbf{71}, 022109 (2005).

\bibitem{Ghosh201286011138} A. Ghosh, S. S. Sinha, and D. S. Ray, Phys. Rev. E \textbf{86}, 011138 (2012).

\bibitem{Ishizaki2008347185} A. Ishizaki and Y. Tanimura, Chem. Phys. \textbf{347}, 185 (2008).

\bibitem{Faddeev1980} L. D. Faddeev and A. A. Slavnov, \textit{Gauge Fields: Introduction to Quantum Theory} (Benjamin-Cummings, Reading, MA, 1980).

\bibitem{Feynman1965} R. P. Feynman and A. R. Hibbs, \textit{Quantum Mechanics and Path
Integrals} (McGraw-Hill, New York, 1965).

\bibitem{Li201081062124} J.-G. Li, J. Zhou, and B. Shao, Phys. Rev. A \textbf{81}, 062124 (2010).

\bibitem{Shen201388033835} H. Z. Shen, M. Qin, and X. X. Yi, Phys. Rev. A \textbf{88}, 033835 (2013).

\bibitem{Shatokhin2000174157} V. N. Shatokhin, S. Ya. Kilin, Opt. Commun. \textbf{174}, 157 (2000).

\bibitem{Lindblad481191976} G. Lindblad, Commun. Math. Phys. \textbf{48}, 119 (1976).

\bibitem{Gorini178211976} V. Gorini, A. Kossakowski, and E. Sudarshan, J. Math. Phys. \textbf{17}, 821 (1976).

\bibitem{Thoss20011152991}M. Thoss, H. Wang, and W. H. Miller, J. Chem. Phys. \textbf{115}, 2991 (2001).

\end{references}
\end{document}